\documentclass[sn-mathphys,Numbered]{sn-jnl}

\usepackage{graphicx}%
\usepackage{multirow}%
\usepackage{amsmath,amssymb,amsfonts}%
\usepackage{amsthm}%
\usepackage{mathrsfs}%
\usepackage[title]{appendix}%
\usepackage{xcolor}%
\usepackage{textcomp}%
\usepackage{manyfoot}%
\usepackage{booktabs}%
\usepackage{algorithm}%
\usepackage{algorithmicx}%
\usepackage{algpseudocode}%
\usepackage{listings}%
\usepackage[multiple]{footmisc}
\usepackage{booktabs}
\usepackage{tablefootnote}

\theoremstyle{thmstyleone}%
\theoremstyle{thmstyletwo}%

\theoremstyle{thmstylethree}%

\begin{document}

\title[]{Analysing and Organising Human Communications for AI Fairness-Related Decisions}
\subtitle{Use Cases from the Public Sector}


\author*[1]{\fnm{Mirthe} \sur{Dankloff}}\email{m.e.dankloff@vu.nl}

\author[2]{\fnm{Vanja} \sur{Skoric}}\email{v.skoric@uva.nl}

\author[2]{\fnm{Giovanni} \sur{Sileno}}\email{g.sileno@uva.nl}

\author[2]{\fnm{Sennay} \sur{Ghebreab}}\email{s.ghebreab@uva.nl}

\author[1]{\fnm{Jacco} \sur{Van Ossenbruggen }}\email{jacco.van.ossenbruggen@vu.nl}

\author*[1]{\fnm{Emma} \sur{Beauxis-Aussalet}}\email{e.m.a.l.beauxisaussalet@vu.nl}

\affil*[1]{\orgdiv{Computer science}, \orgname{Vrije Universiteit Amsterdam}, \orgaddress{\street{De Boelelaan 1105}, \city{Amsterdam}, \postcode{1081 HV Amsterdam}, \country{The Netherlands}}}

\affil[2]{\orgdiv{Informatics Institute}, \orgname{University of Amsterdam}, \orgaddress{\street{Science Park 900}, \city{Amsterdam}, \postcode{1098XH}, \country{The Netherlands}}}


\abstract{AI algorithms used in the public sector, e.g., for allocating social benefits or predicting fraud, often involve multiple public and private stakeholders at various phases of the algorithm's life-cycle. 
Communication issues between these diverse stakeholders can lead to misinterpretation and misuse of algorithms.  
We therefore investigated the communication processes for fairness-related decisions by conducting semi-structured interviews with practitioners working on algorithmic systems in the public sector. 
By applying qualitative coding analysis, we identify key elements of communication processes that underlie fairness-related human decisions. 
We analyze the division of roles and tasks, the required skills, and the challenges perceived by  
stakeholders. 
We formalize the underlying communication issues within a network of stakeholders in a conceptual framework that: 
(i) represents the communication patterns identified in the interviews, and 
(ii) outlines missing elements, such as actors who miss skills or collaborators for their tasks. 
The framework is used for describing and analyzing key organizational issues for fairness-related decisions, for collecting evidence on the communication gaps, and for drafting interventions on the 
patterns of collaboration and communication.
Three general patterns emerge from the resulting analysis: 
(1) Policy-makers, civil servants, and domain experts are less involved compared to developers throughout a system's life-cycle. 
This leads to developers taking on extra roles such as advisor, while they potentially miss the required skills and guidance from domain experts.
(2) End-users and policy-makers often lack the technical skills to interpret a system's limitations and uncertainty, and rely on actors having a developer role for making decisions concerning fairness issues. 
(3) Citizens are structurally absent throughout a system's life-cycle, which may lead to decisions that do not include relevant considerations from impacted stakeholders. 
}


\maketitle

\section{Introduction}\label{Introduction}
Algorithms are increasingly being used for various forms of public sector services such as allocating social benefits in the domains of education, health, and detecting fraud in allowances and taxes \cite{rodolfa, DBLP:conf/fat/RodolfaSHMLG20, williamsoneducation, TNO1, TNO2}. 
These applications can be beneficial, but can also have detrimental consequences for citizens in high-stake scenarios. Notorious examples where incorrect predictions led to wrongful accusations of citizen minorities are the COMPAS case in the US\footnote{Correctional Offender Management Profiling for Alternative Sanctions (COMPAS): the software used to predict the risk of a person recommitting a crime was more inclined to falsely accuse African-American offenders than Caucasian offenders \cite{mehrabi, fassCOMPAS}.}, the SyRI-case\footnote{\raggedright System Risk Indication (SyRI) was a legal instrument used by the Dutch government to detect various forms of fraud, including social benefits, allowances, and taxes fraud. See for instance 'SyRI legislation in breach of European Convention on Human Rights' at \url{https://edu.nl/xjubf}}, and the Childcare Benefit Scandal\footnote{Thousands of families had to repay child welfare subsidies after being wrongly accused of fraud by the tax authority. See for instance the European parliamentary questions at \url{https://edu.nl/y3h3j}.} in the Netherlands. The latter eventually led to the resignation of the Dutch government in 2021\footnote{See e.g. ``Dutch Government resigns over Child Benefit Scandal'', The Guardian (1, 2021),\url{https://www.theguardian.com/world/2021/jan/15/dutch-government-resigns-over-child-benefits-scandal}}.

These examples highlight the problem of fairness in AI. \textit{Fairness} in this context refers to \textit{fair outcomes} for decision-making, 
a principle that prescribes that algorithmic decision-making must have an absence of prejudice or favoritism toward an individual or group based on their inherent or acquired characteristics \cite{mehrabi}.
Nowadays, fairness --and related issues in AI -- are widely recognized in well-established legal and ethical guidelines \cite{EU1, EU2, AIACT}. 
According to the European Commission's Ethics guidelines on trustworthy AI \cite{EU2}, an important step in supporting trustworthy AI includes involving and educating all stakeholders about their roles and needs throughout the AI system's life-cycle. 
Indeed, algorithms are always part of a process driven by many stakeholders' design choices and socio-cultural norms
\cite{Suresh}. All the (design) decisions that are made throughout a system's life-cycle codify the underlying socio-cultural norms of the stakeholders \cite{lee, amershi, haakman}. 
For instance, when allocating benefits in the public sector, it has to be decided which data features are relevant for the `eligibility' for social benefits \cite{barocas-hardt-narayanan}. Furthermore, the punitive (e.g., detecting fraudsters) or assistive (e.g., allocating social benefits) nature of policy interventions might require balancing false positive and false negative rates \cite{Saleiro,DBLP:conf/fat/RodolfaSHMLG20}. 
Therefore, a solely technical approach to fairness is insufficient, and involving diverse actors and stakeholders is important for ensuring that public interests are prioritized and that potential harms are minimized \cite{stapleton, filgueiras}.

To address such issues, we investigate the communication and collaborations between stakeholders throughout an algorithm's life-cycle. In this paper, we use a working definition of \textit{fairness-related decisions} for all design decisions and practices applied by stakeholders that can potentially lead to bias, discrimination, and other forms of prejudice against different groups, individuals, or communities \cite{Suresh, barocas-hardt-narayanan, Madaio}.
We do not consider a predefined scope of \textit{fairness-related decisions }and focus on the non-exhaustive scope of decisions that emerge along our investigations.
We focus on internal communications between the direct stakeholders who use or build a system, rather than external communications with the general public \cite{Madaio, FestWieringa}. We do this by identifying the roles, the divisions of tasks, the required skills, and the potential communication challenges between diverse actors occurring throughout the algorithm's life-cycle. The research questions we address are the following: 

\begin{itemize}
\item \textbf{RQ1:} Which actors, roles, and tasks can be identified in multi-stakeholder interactions throughout the phases of an algorithm's life-cycle when making fairness-related decisions?
\item \textbf{RQ2:} Which communication patterns and challenges can be identified when stakeholders make fairness-related decisions? 
\end{itemize}

To answer these questions, we conducted 11 semi-structured in-depth interviews with public practitioners working on algorithmic systems. 
For reasons of better accessibility, we concentrated on experts from organizations in the Netherlands, but the methodology applied in the study can be easily replicated in other contexts to further extend our results. 
From the interviews, we identified who makes decisions about what, and at which phase of the algorithm's life-cycle. 
We analyzed the interview transcripts to identify the elements that constitute communication patterns and challenges, and we labeled them through \textit{in-vivo}, descriptive, and process coding \cite{Saldana}. 

We further structured our findings by building a conceptual framework that draws the key relationships between the constitutive elements of communication patterns that underlie fairness-related human decisions. 
First, we found that it is crucial to differentiate stakeholders by the individual \textit{actors}, the \textit{roles} that actors assume when contributing to a \textit{task} that involves fairness-related decisions, and the \textit{skills} that a task requires or an actor has.
For example, simply describing a stakeholder as a developer can omit to indicate that the same actor (i.e., the same person) also assumes the role of advisor with domain expertise when they decide which features are to be used as predictors for detecting fraud. 
Not only do such actors endorse more than a developer role, but they can also miss the skills required for their extra role.
Second, we found that it is crucial to identify the elements that stakeholders are missing, to describe the communication challenges that stakeholders experience when making fairness-related decisions.
Thus, the conceptual framework we derived for analyzing the communication patterns in fairness-related decisions has 3 main characteristics: (1) it differentiates actors from their roles or skills; (2) it considers 6 key elements of communication patterns: Actors, Roles, Skills, Tasks, Information exchange, and Phases in the algorithm's life-cycle; and (3) it can specify the elements that are deemed missing in the communication patterns.
After analyzing the interview transcripts using this conceptual framework, we formalized 3 general patterns that emerged from the participants:
(1) Developers play the most prominent role in most tasks and phases of the algorithm’s life-cycle even though they miss guidance from stakeholders with advisor and policy-maker roles and domain expertise skills; 
(2) end-users and policy-makers often lack the technical skills to interpret a system's limitations and uncertainty, and the related fairness implications; and (3) inputs from citizens are structurally absent in fairness-related decisions throughout a algorithm's life-cycle.

These communication challenges indicate inadequate model governance, and the potential inability to recognize and address fairness issues throughout an algorithm's life-cycle. 
This can lead to misinterpretation and misuse of algorithms, with critical implications for the impacted populations. 
The communication challenges we identified, and the conceptual framework we derived may help identify such issues before they arise in practice, after algorithms are deployed.  

\section{Related Work}\label{Related Work}
Several frameworks and theories from various domains have been proposed to characterize the dynamics of interactions amongst a network of actors \cite{Ropohl1999, Latour1999}.
Actor-Network Theory (ANT) and mediation theory, for example, describe the relations and interactions within a network of (artificial and natural) actors \cite{latour1992, latour1994, Latour1999}. 
Following ANT, interaction with technology is never neutral as it influences or mediates the way tasks and decisions are carried out. On the other hand, technology is continuously mediated by human social aspects, e.g. in formulating design goals. To describe the context of reciprocal interactions between human actors and technology, we can broadly refer to socio-technical systems (STS) approaches \cite{Ropohl1999}. A view centered on STS does not consider technology alone, it rather stresses the interactive nature of social and technical structures within an organization or society as a whole. This approach is increasingly used in the field of AI, to assess fairness and ethics from a broader normative context in which actors interact and operate, as opposed to focusing on individual actors alone \cite{Chopra, dolataSTS, slota}. 

Other frameworks have been proposed to investigate the power structures within a network of actors. Following the tripartite model for ethics in technology, three main roles are often identified through their responsibilities: (1) the developer, who handles the technical aspects; (2) the user, who handles the practical usage of the system, and (3) the regulator's role, who is responsible for making the value decisions \cite{Poel}. Prior research on automated systems for public decision-making has shown a shift of discretionary power from the regulator roles to developer roles, often making the latter the main decision-makers \cite{Bovens2}. When developers become the main decision-makers for design decisions, this can exclude stakeholders without technical knowledge from important decisions about the system \cite{Kalluri, Danaher}. These imbalanced power dynamics can lead to a form of technocracy, where governance and (moral) decision-making are based on technological insights and may only yield technological solutions \cite{Poel, hickok, filgueiras}.

Beyond these theoretic considerations, empirical field research has been conducted to investigate data practices at local governments \cite{Siffels, FestWieringa, Jonk}. For instance, Siffels et al. (2022) argue that with the process of decentralization in the Netherlands, many tasks from the central government were delegated to municipalities without giving them more resources and capacities. Municipalities invested in data practices to deal with additional tasks and to distribute limited (social) resources. Due to a lack of data literacy, however, public servants were unable to recognize ethical issues and thus sought collaboration with external partners. Other research showed that depending on their roles and tasks, stakeholders can be involved at different phases in the algorithm's life-cycle \cite{Wieringa, Bovens, TNO2, Spierings}. Decision-makers from public organizations are often involved in the procurement and deployment phases. Developers, sometimes from third parties, tend to be more involved in the development phase \cite{Wieringa, Spierings}. This can sometimes lead to ``The problem of many hands", which indicates a decreased ability to be transparent and responsible, because parts of the management of the algorithm's life-cycle are outsourced to different stakeholders \cite{Siffels, cobbe}. 
Jonk and Iren (2021) performed semi-structured interviews with practitioners at 8 municipalities, to investigate the actual and intended use of algorithms \cite{Jonk}. They found a lack of common terminology and algorithmic expertise, at a technical level and at a governance and operational level. The authors argue that municipalities would benefit from a governance framework to guide them in the use of tools, methods, and good practices to handle potential risks. Lastly, Fest, Wieringa, and Wagner (2022) investigated how higher-level ethical and legal frameworks influence daily practices for data and algorithms used in the Dutch public sector \cite{FestWieringa}. They found that applying existing frameworks remains challenging for practitioners because they do not feel competent or miss the required skills to make decisions for their practices to be responsible and accountable. Data professionals, as a result, get too much autonomy and discretion power for handling decisions that belong to the core of public sector operations and mandates. 

What is still missing in previous work is a framework to characterize the communication processes that underlie fairness-related human decisions throughout an algorithm’s life-cycle. The frameworks and theories in related works indicate that such communication and decision processes arise within a socio-technical interactive network, where algorithms are part of a governance structure comprising actors with different roles and tasks. The literature also shows that our research must consider the interactions between stakeholders who have direct or indirect interactions with an algorithm, and with the populations impacted by the algorithm. Thus, we aim at identifying how fairness-related decisions are mediated by stakeholders who may or may not have direct access to socio-technical information that is relevant for addressing fairness issues.

\section{Methodology}\label{Methodology}
\subsection{Semi-structured interviews}\label{subsec2}
We conducted 11 semi-structured interviews. Each interview lasted for approximately one hour. We formulated the interview questions in an open-ended manner, where participants were able to share their information in their own words whilst following a general structure of topics \cite{Interview1, Interview2}. Before conducting the interviews, participants received some example questions and a short description of the research. At the start of the interview, participants gave their consent for their interview to be used in this research. Also, they were asked to discuss one use case they were involved in. The questions used for the interviews can be found in Table \ref{tab:structured_interviews} in the appendix and are divided into three main sections:
\begin{enumerate}
    \item [1] \textbf{General}: Investigation of the project and use case to which the participant contributed, the other actors involved, and the participant's team, roles, and envisioned (end) users. 
    \item [2] \textbf{Development process}: Investigation of the type of datasets, resources, tasks, phases, and roles needed throughout the algorithm's life-cycle to make fairness-related decisions. 
    \item [3] \textbf{Considerations}: Investigation of the perceived challenges for role and task division, the potential improvements or failures of the system, and the communication gaps. The questions also concerned the assessment of error and bias, and the the potential negative impacts of the algorithm.
\end{enumerate}
In the first two sections, participants were asked to describe the general procedures and practices used in the AI system's life-cycle. Participants had the opportunity to mention internal communication and key elements of the communication processes that underlie fairness-related human decisions. We specifically asked about communication issues in the third section of the interview. This division was made to provide the opportunity for spontaneous answers beyond our specific questions.

We preliminary tested all interview questions with a pilot with 5 researchers from different disciplines in our research lab. The questions were deemed suitable for letting participants describe their communication process and related issues. The suitability of the questions was checked in terms of comprehensibility and relevance to our research questions. No questions were altered afterwards.

\subsection{Case Studies}
We recruited participants who have been collaborating on multi-stakeholder projects in the public sector. Participants working in the social domain, e.g. social benefit allocation or fraud detection were of particular interest because the impacts on citizens can be critical.
We used a repository of use cases that was made available to us by the Dutch Ministry of Interior Affairs\footnote{Some examples of public domain use cases in the Netherlands can also be found via the Artificial Intelligence Netherlands Coalition (NL AI Coalitie) website (\url{https://nlaic.com/use-cases}) and in \cite{TNO1,TNO2}.}$^,$\footnote{Dutch Ministry of Interior Affairs and Kingdom Relations \url{https://www.rijksoverheid.nl/ministeries/ministerie-van-binnenlandse-zaken-en-koninkrijksrelaties}}. Next to that, we used the \textit{snowball sampling} technique to recruit participants. 

\begin{table}[!t]
\centering
  \begin{tabular}{lll}
    \toprule
    Interviewee & Role & Technical background \\
    \midrule
    P1 & Developer \& Researcher & yes \\
    P2 & Manager \& Researcher & no \\
    P3 & Manager & no  \\
    P4 & Manager & yes \\
    P5 & Advisor \& Researcher & no  \\
    P6 & Developer \& Researcher & yes\\
    P7 & Manager & no  \\
    P8 & Advisor \& Researcher & yes \\
    P9 & Developer \& Researcher & yes  \\
    P10 & Advisor \& Researcher & no \\
    P11 & Manager & yes \\
  \bottomrule
\end{tabular}

\caption{Description of participants}
\label{tab:participants}
\end{table}

Table \ref{tab:participants} describes the participants, their roles at the time of involvement, and if they have a technical background. We consider those who are not educated or have no experience in technical science to not have a technical background. 10 participants were involved in the social security domain, and 1 participant was in the education domain. 

\subsection{Qualitative Coding analysis}
We performed a qualitative coding analysis by labeling key \textit{codes}\footnote{Saldaña (2013) describes that ``A code in qualitative inquiry is most often a word or short phrase that symbolically assigns a summative, salient, essence-capturing, and/or evocative attribute for a portion of language-based or visual data." In addition, a code can be understood as a researcher-generated construct that symbolizes the construct and assigns an interpreted meaning.} from the interview output. We used \textit{in vivo}\footnote{``In Vivo'' coding is also named ``literal coding'' and refers to a word or short phrase from the actual language found in the qualitative data record e.g., terms used by participants themselves \cite{strauss, Saldana}.}, descriptive\footnote{Descriptive coding refers to summarizing the basic topic of a passage of qualitative coding in a word (noun) or short phrase \cite{Saldana}.} and process coding\footnote{Process coding refers to ``action coding'' which implies action from more simple observable activity (e.g.,  reading) to more general conceptual action (such as adapting) \cite{Saldana}.} to identify the process of communication exchange between diverse actors, as well as the practices and choices made at each stage of the algorithm's life-cycle. 

The coding analysis was performed in multiple cycles. At each round of coding, pieces of text are annotated with \textit{codes} that represent the concepts mentioned by participants. The codes are refined, merged, or split into categories after each round. This was repeated until no further refinement of the codes was needed. Two of the authors perform a separate coding analysis to reduce the impact of personal bias. We performed coding analysis by hand and using a coding analysis tool\footnote{Atlas.ti: The Qualitative Data Analysis \& Research Software \url{https://atlasti.com/}}. We compared both coding analyses to identify discrepancies or alignments. Beforehand, both analysts agreed that particular attention should be drawn to identifying the roles, tasks, phases, and challenges from the interview transcripts. For example, if a participant were to mention that "[person X] is a developer and performs bias analysis in the development phase", the actor, the role, the task, and the phase would be labeled.

\begin{figure}[t!]
    \centering
    \includegraphics[width=\linewidth] {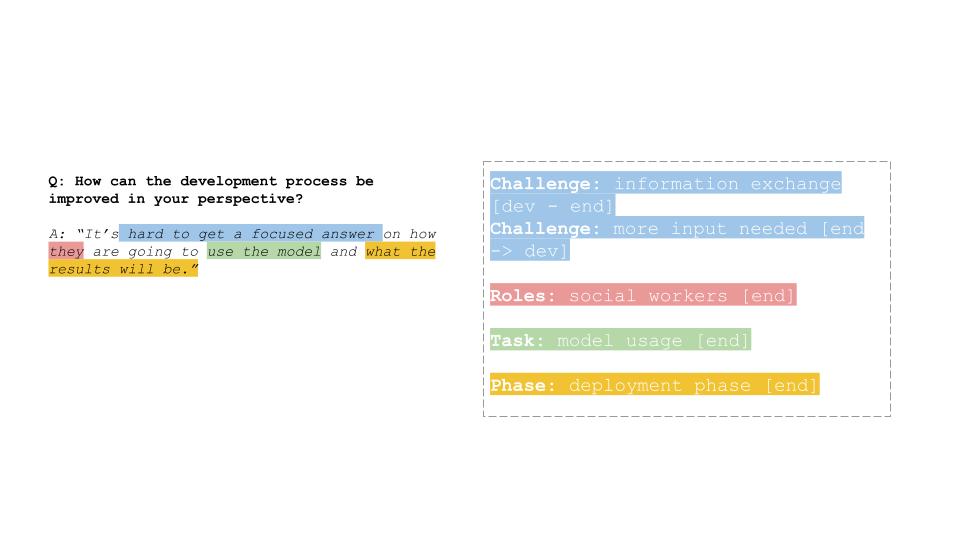} \vspace{-50pt}
    \caption{Example of interview Q\&A (left) and the corresponding coding labels (right) from the qualitative coding analysis}
    \label{fig:coding}
\end{figure}

In Figure \ref{fig:coding}, an example is given for the interview output (left) and the corresponding codes (right). The Figure shows the colors corresponding to the groups of codes for \textit{challenge, roles, task}, and \textit{phase}. On the right, an example of the corresponding descriptive codes can be found. For example, ``it's hard to get a focused answer" was summarized as an information exchange challenge of the type where ``more input is needed". We added the corresponding role(s) to the codes in brackets ``[]". If the code concerned multiple roles, we added ``-" to indicate a relation for information exchange. In this example, more input is needed between the end-user and the developer role. 

After all interview transcripts were annotated with codes, we analyzed which codes co-occurred within the answers to each question, e.g. we counted which roles occurred together with a specific phase, task, or challenge.

\subsection{Constructing a conceptual framework}
The co-occurrence analysis alone did not capture the relations between codes, i.e. ``end-user is missing in the development phase" would still count as a co-occurrence of the codes \textit{end-user} (role) and \textit{development} (phase), although the role of end-user was actually missing. Therefore, we constructed a framework that further analyses the codes we identified by describing their relations and characteristics. The conceptual framework aims to describe the key codes and relationships between high-level groups of codes (e.g. Actor, Role, Skill, Task, Phases), and the key characteristic that underlie the challenges mentioned in the interviews. We can then represent challenges such as citizens are actors with the role of \textit{Data subject} (an Actor-Role relationship), and that actors with such roles are ``missing" (a characteristic of Actors). 

We constructed the conceptual framework iteratively, following a method similar to those used for constructing ontologies \cite{Noy, SEM, Golpayegani, FMOontology}. This means that we continuously adjusted the framework until it would represent every code we identified from the qualitative coding analysis. We added definitions, characteristics, and properties to the identified concepts. 
We added descriptions to each concept to agree on common definitions. The relationships and characteristics we used to build the conceptual framework are based on the interviews and were in accordance with some of the definitions we found from documents provided by the European Commission on Trustworthy AI, and from other sources in the literature \cite{EU1, EU2, SEM, Golpayegani, TamburriSkill, Poel}. For example, by describing the type of private or public affiliation (e.g., national institute, ministry, or municipality) we can contextualize how tasks and roles are divided within multi-stakeholder collaborations. Relations are added between codes. For example, an actor always ``has" a certain role whereas a task ``involves" a role ``during" a phase. 

\section{Results}\label{sec6}
In the next section, we first describe the use cases discussed by the participants in the interviews ({section~\ref{sec_cases}}). Then the results of our qualitative coding analysis are given in {section~\ref{sec_coding}}. Finally, we further analyze the communication challenges and apply them to document the communication patterns and challenges we identified in {Section~\ref{sec_parts}}.

\subsection{Use cases}\label{sec_cases}
In all use cases, multiple stakeholders were involved with varying expertise---from social workers to developers, researchers, program managers, and advisors from third parties. For most use cases (10 out of 11), the procurement for the algorithm came from government organizations and municipalities. Furthermore, the envisioned end-users of the systems were in 10 out of 11 cases policy-makers or social workers at municipalities with minimal or no technical expertise. End-users and policy-makers were mentioned to be the same in most of our use cases. For the remaining use case in the educational domain, teachers were the envisioned end-users.

\subsection{Qualitative Coding Analysis}\label{sec_coding}
\subsubsection{Identified Codes and Concepts}
Our qualitative coding analysis first focused on identifying the main types of Roles, Tasks, and Challenges. It resulted in identifying 7 codes for describing the main roles (Table \ref{tab:main_roles}), 10 codes for the tasks (Table \ref{tab:main_tasks}), and 7 codes for the challenges (Table \ref{tab:main_challenges}.) In this section, we explain in more detail the concepts that these codes represent, and our decisions for eliciting a consistent set of codes.

For coding the roles, we observed that the terminology is rather diverse for the technical roles.  
For example, participants mentioned terms such as engineers, coders, developers, and data scientists for the role of developer. 

Some participants identified themselves or their collaborators as researchers. We questioned the inclusion of code for the role of researcher. However, such code can be ambiguous as the research topics could either concern the technical development of algorithms, or other domains such as governance or social security.
Thus, we decided to group under the code ``developer" the researchers who focus on the technical development of algorithms. Researchers that contributed from other domains sometimes assumed roles other than developers, such as advisor or manager.

For the role of manager, participants mentioned terms such as innovation managers, product owners, program managers, project managers, or CTO (Chief Technology Officer). These terms were often used interchangeably. We decided to group all management-related roles under the same code ``manager'', without using specific codes for each job title or hierarchical level.

In Table \ref{tab:main_roles} the descriptions of the main roles can be found. For example, managers are, respectively, those who ``supervise the projects for the development of the system and oversee documentation checks and balances''. In our use cases, the managers often worked in the same team as developers and were either hired externally or internally by a (public) requester. 

The request for the model ---associated with the ``requester'' role--- often came from ministries, and they were only mentioned for funding or initiating a project.

\begin{figure}[t]
\centering
\scalebox{0.45}{
\includegraphics{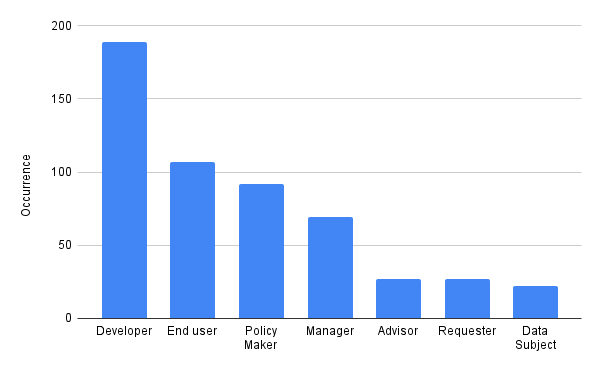}
}
\caption{Main Roles Identified. The number of times a Role is mentioned. Note that Developers are mentioned most \textit{(N = 189) }and Data Subjects least \textit{(N = 22).}}
\label{fig:mainRoles}
\end{figure}

The ``data subject'' role as well as the ``requester'' role were never described as the end-users. In Table \ref{tab:main_roles} it is also stated that data subjects are ``an organization or entity that is impacted by the system, service or product'' \cite{Golpayegani}. The data subjects were, in almost all of our use cases (10 out of 11), citizens.
The advisor role was often presented as advising on 1) domain knowledge, 2) technical knowledge, or 3) ethical knowledge.
Overall, as illustrated in Fig.~\ref{fig:mainRoles}, the developer role was mentioned the most \textit{(N=189)}, followed by End-users \textit{(N=107)} and Policy Makers \textit{(N=92)}. 

In Section \ref{co-occurences}, we describe which roles occurred the most for which phase (\ref{Roles and Phases}) of the algorithm's life-cycle. 

\begin{table}[t!]
\centering
  \begin{tabular}{lp{0.7\linewidth}}
    \toprule
    Role types & Description \\
    \midrule
    Developer &  Research, design, and/or develop algorithms \\
    Policy-maker & Responsible for designing and overseeing the carrying out of policy and social decisions \\
    Manager & Supervise the projects for the development of the system and oversee documentation checks and balances \\
    End-user & (In)directly engage with the system and use algorithms within their business processes to offer products and services to others \\
    Data subject & Organization or entity that is impacted by the system, service, or product  \\
    Advisor & Give constructive feedback on the system throughout the life-cycle \\
    Requester & Who are the main client and investor for the use-case \\
    \bottomrule
\end{tabular}
\caption{Descriptions of Main Roles }
  \label{tab:main_roles}
\end{table}

\begin{table}
\centering
  \begin{tabular}{lp{0.6\linewidth}}
    \toprule
    Task& Description \\
    \midrule
   Technical Decision& Decision-making for technical aspects in the development, training, and testing of the AI model\\
   Consulting & Advising on domain, technical, or ethical knowledge aspects of the AI model\\
   Fairness \& Risks & Controlling the negative social impacts of the AI model such as discrimination\\
   Involving Stakeholders & Actively involving other actors throughout the AI algorithm's life-cycle (e.g., a task of manager roles)\\
   Researching & Examining, studying or investigating aspects of the AI algorithms' life-cycle\\
   Goal Formulation & Planning and deciding on the purpose of the AI model and its requirements\\
   Bias Analysis & Analysing the systematic differences between populations or individuals in the AI model output\\
   Model Usage& Operationalizing the AI model\\Go - no go& Deciding to proceed with the development and or implementation of the model\\
   Auditing & Documenting and controlling the process of and around the AI model\\
   \bottomrule
\end{tabular}
\caption{Descriptions of Main Tasks}
  \label{tab:main_tasks}
\end{table}

In Table \ref{tab:main_tasks}, we describe the main tasks identified from the qualitative coding analysis. For example,  the task "Consulting" refers to advising on domain, technical, or ethical knowledge aspects of the AI model. In Section \ref{Roles and Tasks}, we describe which roles occurred the most for which task. 

\begin{table}[t!]
  \begin{tabular}{lp{0.7\linewidth}}
    \toprule
    Challenge & Description \\
    \midrule
    Interpretation & Misunderstanding of technical information regarding the AI model, such as evaluation metrics\\
    Involvement & Lack of participation and collaboration between actors throughout the algorithm's life-cycle \\
    Risk Oversight & Problems concerning governance, legal, ethical, and procedural aspects  \\
    Resources & Insufficient time, planning, infrastructure, money, and documentation \\
    Feedback & Lack of substantial input, information exchange between actors  \\
    Bias & Problems with the analysis of prejudice towards individuals or groups  \\
    Role & Unclear function or duty division among actors\\
  \bottomrule
\end{tabular}
\caption{Descriptions of Main Challenges}
  \label{tab:main_challenges}\vspace{-10pt}
\end{table}

In Table \ref{tab:main_challenges}, we describe the main challenges identified from the qualitative coding analysis. For example, "Interpretation" issues refer to the misunderstanding or misevaluation of information regarding the AI model. 
In Section \ref{Roles and Challenges}, we describe which roles occurred the most for which challenge. 

\subsubsection{Co-occurrences}\label{co-occurences}

\begin{figure}[t!]
\centering
\scalebox{0.4}{
\includegraphics{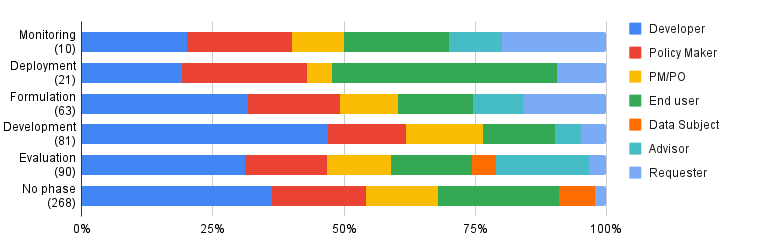}
}\vspace{5pt}
\caption{Co-occurrences of Roles and Phases. The number of times a Phase is mentioned is shown on the y-axis (in decreasing order). Note that the developer role (blue) is mentioned the most in all phases and that the data subject role (orange) is mentioned the least.}
\label{fig:mainrolexphase}
\end{figure}

\begin{figure}[t!]
\centering
\scalebox{0.4}{
\includegraphics{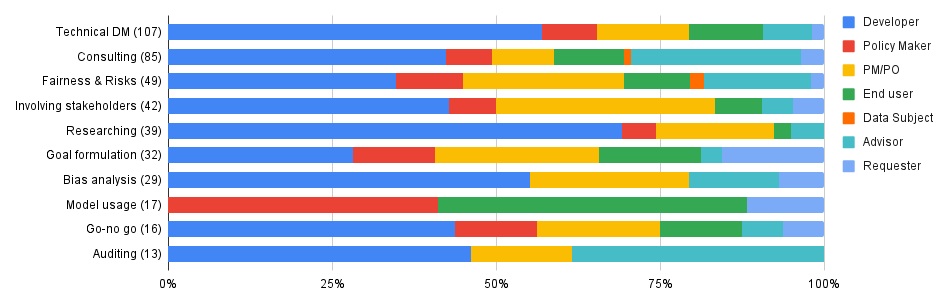}}\vspace{-10pt}
\caption{Co-occurrences of Roles and Tasks. The number of times a Task is mentioned is shown on the y-axis (in increasing order). Note that technical decision-making is mentioned the most. Developer roles (blue) are mentioned the most for all tasks except for model usage.}
\label{fig:mainrolextask}
\end{figure}

\begin{figure}[t!]
\centering
\scalebox{0.4}{
\includegraphics{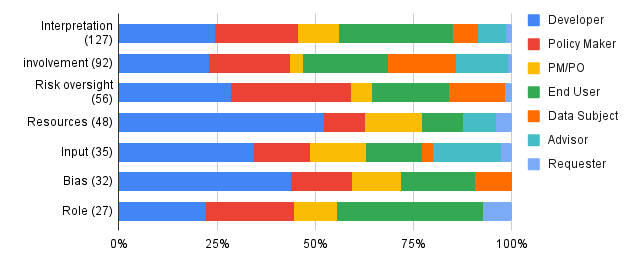}}
\caption{Co-occurrences of Roles and Challenges.  The number of times a Challenge is mentioned is shown on the y-axis (in increasing order). The roles of developer (blue), end-users (green), and policy-makers (red) are mentioned most.}
\label{fig:mainrolexchallenge}
\end{figure}

Figures~\ref{fig:mainrolexphase},~\ref{fig:mainrolextask},~and~\ref{fig:mainrolexchallenge} illustrate which Roles were mentioned the most, based on co-occurrence with Phases (Section \ref{Roles and Phases}), Tasks (Section \ref{Roles and Tasks}), and Challenges (Section \ref{Roles and Challenges}). 

\paragraph{Roles and Phases}\label{Roles and Phases}

Figure \ref{fig:mainrolexphase} shows that \textbf{developers are most prominent in the development, evaluation, and formulation phases, but less prominent in the deployment and monitoring phases}. Developer (P1) mentioned that \textit{``we don't monitor what the municipalities are doing with the results.''} and \textit{``feedback is needed on how the results will be used in deployment''.} 
\textbf{Conversely, stakeholders other than developers could be more involved in the development phase}. Another developer (P9) mentioned \textit{``For the future, we could incorporate stakeholders at earlier stages in the development to see what the potential sources of bias are.''} 

End-users and policy-makers were the second highest in occurrences for phases. Moreover, Figure \ref{fig:mainrolexphase} demonstrates that the monitoring phase \textit{(N = 10)} was mentioned the least throughout the interviews whereas the evaluation phase was mentioned the most \textit{(N = 90)}. 

Data subjects were seldom mentioned to be involved. \textbf{Data subjects could be more involved throughout the phases of an algorithm's life-cycle} e.g. P5 mentioned \textit{``it depends on the type of AI. If it has an impact on citizens or uses a lot of data from citizens, it would be relevant to include a focus group of citizens from the beginning but it is less relevant for road repairs.''}. The role of the requester was only mentioned in the formulation phases but rarely as being involved throughout other phases. Advisor roles were often mentioned to be involved in the evaluation phase before deployment, or when the project is halted.

\paragraph{Roles and Tasks}\label{Roles and Tasks}

Figure \ref{fig:mainrolextask} shows that the developer role was mentioned the most for all tasks (e.g., technical decision-making, consulting, dealing with fairness and risks). \textbf{This indicates that actors taking on developer roles were the most prominent in making decisions throughout the algorithm's life-cycle.} About the typical tasks developers handle, developer (P9) mentioned that they \textit{``decided on how to improve accuracy and handling issues. For instance, gathering more diverse data to handle bias''}. About their collaboration with other roles, another developer (P1) mentioned that they \textit{``define and chose metrics for the models''} and that these \textit{"are defined in collaboration with the municipality but choosing metrics and trimming down after input was decided by the two of their team''}. The developer role was not mentioned for tasks related to model usage.

Regarding the task of stakeholder involvement, managers are the main decision-makers. Within teams, managers are sometimes the only ones in direct contact with roles other than developers. Managers were often mentioned to supervise developers in technical decision-making, and they often rely on the developers' judgment for bias and risk oversight. A manager we interviewed (P2) mentioned that for handling error rates and biases they \textit{``rely on the technical teams' judgment''} and that \textit{``the technical colleagues give advice when the model is good enough, but it's a bit of a grey area. We also rely on literature''.} Another manager (P4) mentioned that \textit{``it is time intensive to explain [bias analysis] to stakeholder users. Bias analysis is sometimes so complex, even as an expert I sometimes don't understand it, and it takes a lot of time''.} 

Actors with a developer role also sometimes assume advisor roles. When technical advisors are missing, managers can hire a third-party developer to analyze the code, give technical advice, or even build the model. 
An advisor we interviewed (P5) mentioned that \textit{``an external company was hired to develop the model for the municipality''}, which made the \textit{``data ecosystem quite complex''}. 
Another manager we interviewed (P3) added that they \textit{``hired an external bureau for auditing and investigating the algorithm''}, e.g., as they \textit{``could not get reliable predictions because the social domain changes all the time, and it's hard to keep track of these changes---for example in social support---and how that impacts the system''}. 

Advisor (P10) mentioned that they \textit{``were involved to give feedback as an involved bystander. But it was hard for someone like me to understand what the difference between implementation and design is and what that means for real-life implications''.} This demonstrates that \textbf{roles other than developers lack the technical skills to participate in decisions}.

\paragraph{Roles and Challenges} \label{Roles and Challenges}

In Figure \ref{fig:mainrolexchallenge} the co-occurrences of roles and challenges are shown. Most communication challenges were associated with the roles of developers, end-users, and policy-makers.

The role of the end-user was frequently mentioned for challenges related to interpretation and role. This means that most challenges were related to either the (mis)understanding or (mis)evaluation of information or an unclear function or duty division amongst actors. Several participants mentioned that more input is needed from end-users on the interpretation and use of the results envisioned in the deployment phase. For instance, a manager (P3) mentioned that it is challenging that \textit{``we don't know if governments and municipalities can understand the model''}. A developer (P1) also mentioned that \textit{``it's hard to get a focused answer on how they are going to use the model and what the results will be'}, and that \textit{``the municipality is too loosely involved in the project.}

Regarding the challenges with bias, risk oversight, and the interpretation of model output, the role of policy maker was frequently mentioned for the challenges related to risk oversight. This concerns problems in governance, legal, ethical, and procedural aspects. The role of developer was frequently mentioned for challenges related to resources, input, and bias. It was mentioned that more input is needed on the analysis of feature selection and bias in the development phase from developers to end-users and policy makers. \textbf{Both end-users and policy-makers were often mentioned as missing the technical skills to understand the uncertainty of predictions and limitations of the model in real-world settings.} A developer (P6) mentioned about end-users that \textit{``people could trust the model blindly and mistake it for a decision-making tool''} and another advisor (P8) where inspectors were the end users mentioned that they were \textit{``not sure if the inspectors fully understood why certain cases were flagged as misuses or put on the list} [of potential frauds]''. 

With regard to challenges for bias and risk oversight, an advisor (P8) mentioned that \textit{``there should be more focus on asking users what policy-makers perceive as risks and biases''} and that it is \textit{``difficult for them to understand that there are many different interpretations. What it really means to be a 'true positive', is this person really a fraud, or was this person not able to fill in the forms properly?''}. 
Interpretation challenges by end-users and policy-makers were mentioned most in the monitoring and deployment phases. A developer (P6) mentioned that \textit{``training for users is needed, to remind users not to rely on the tool but that the decision is up to them.''}. 

Data subjects (e.g. citizens) were also mentioned for involvement and risk oversight challenges. Participants mentioned a \textbf{ need for more citizen involvement and for being more transparent to citizens throughout the phases of an algorithm's life-cycle. } Managers are looking for appropriate frameworks for (fruitful) collaborations with citizens. Manager (P4) mentioned that ``\textit{there is a long history with the citizen council for consultation and it is usually conflict-based. It's hard to make fruitful collaboration, getting them to understand the issues and getting them out of anger mode''}. An advisor (P7) mentioned about previous involvement with citizens that \textit{``they} [citizens] \textit{said no on the feasibility of the model from the municipality. They did not get it. It was more of a general no to technology instead of asking a targeted question''. } 

\subsubsection{Key Insights from Qualitative Coding Analysis}
From the qualitative coding analysis, we conclude that: (1) Actors with developer roles are predominant in most phases and tasks while potentially lacking the required guidance from domain experts. (2) End-users and policy makers often lack the technical skills to interpret a systems output or estimate potential fairness issues. (3) Citizens filling the role of data subjects are seldom mentioned to be involved throughout the phases of the algorithm's life-cycle. In the next section, we analyze these challenges further by characterizing the relations between the main elements of communication patterns in a conceptual framework. 

\subsection{Modeling Communication Patterns 
}\label{sec_parts}
\begin{figure}[t!]
\centering
\includegraphics[width= 8cm]{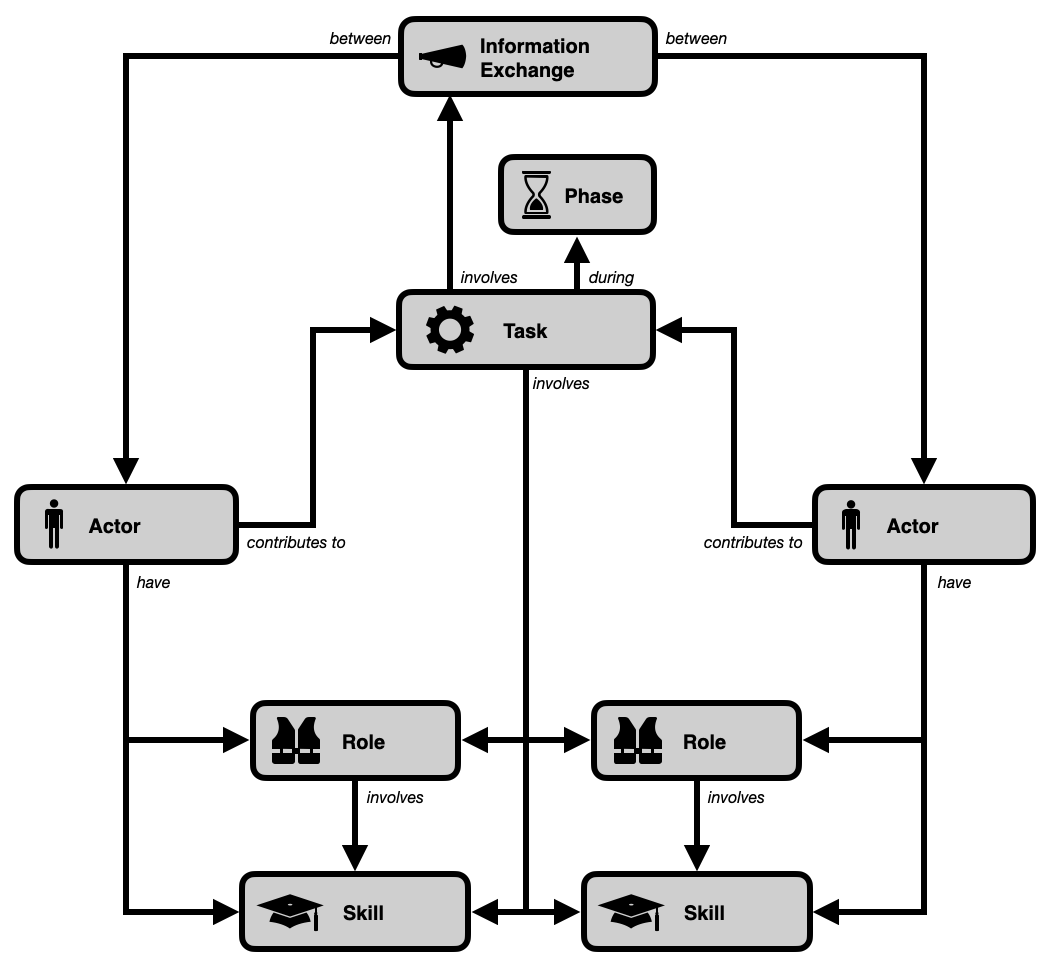}\vspace{5pt}
\caption{The basic concepts selected from our qualitative coding analysis, and used to characterize the communication patterns underlying fairness-related decisions, and the challenges we identified. } 
\label{fig:parts_model}
\end{figure}

The communication patterns emerging from the interviews are not easily described with qualitative analysis in written form only. The codes may identify the key elements of communication patterns, but not their relationships. Counting the (co-)occurrences of codes could not fully capture these relationships. We thus elicited a conceptual framework that models the relationships between the elements of the communication patterns (e.g., between actors, roles, and skills). This also allowed us to explore the perspective of socio-technical systems (Section \ref{Related Work}), in which AI models and fairness-related decisions arise through the interactions between actors. The conceptual framework we elicited is shown in Figure \ref{fig:parts_model} and detailed in Table \ref{tab:concept_descriptions}.

We elicited 6 elements to describe the communication patterns: Phase, Role, Task, Skill, Actor, and Information Exchange. 
At least two concepts were needed to characterize the communication process: the stakeholders who exchange information (represented by the concept \textit{Actor}), and the act of communicating (represented by the concept \textit{Information Exchange}). 
Describing the context of the communications requires at least 4 additional concepts (\textit{Phase, Task, Role, Skill}) to underlie fairness-related human decision-making and their challenges. 
For example, \textit{Tasks} may be missing at certain \textit{Phases} of a system's life-cycle. Or \textit{Actors} may not have the right \textit{Skill} or \textit{Role} when making a fairness-related decision. 
\textit{Skill} was added as a key element of the communication pattern because the interviews showed that the challenges that stakeholders face often arise from the mismatch between their \textit{role} and \textit{skill}.

\begin{table}[t!]
  \centering

  \begin{tabular}{p{0.12\linewidth}p{0.26\linewidth}p{0.26\linewidth}p{0.22\linewidth}}
    \toprule
    Concept & Description & Relation & Property \\
    \midrule
    Task & Actions that Actors perform & Involves Role, Skill, Info exchange, During Phase, Actors contribute to & \textit{is missing} \\
    
    Role & Function filled by Actor & Actors Have Role, Task involves Role, Role involves Skill & \textit{is missing} \\
    
    Actor & Entities that perform Tasks, actively or passively & have Role, Skill, Contributes to Task, Info Exchange between Actors & \textit{affiliation}: public (national institution, ministry, municipality) or private, \textit{is missing} \\
    
    Phase & Indicates the evolution of the system from conception through retirement & Task during Phase & \textit{is missing} \\
    
    Skill & Professional ability, expertise, or knowledge needed in practice to complete a specific task & Task involves Skill, Actor(s) have Skill, Role involves Skill & \textit{is missing} \\
    
    Information exchange & Communication transfer between Actors & between Actor(s), Task involves Info exchange & \textit{is missing} \\
  \bottomrule
\end{tabular}
\vspace{5pt}
  \caption{Description of concepts used to characterize the communication patterns underlying fairness-related decisions}
  \label{tab:concept_descriptions}
\end{table} 

To provide a temporal overview of the communication processes, we link the \textit{Tasks} to the \textit{Phases} of the system's life-cycle in which they take place 
(e.g., to reflect on the fairness-related) \textit{Tasks} 
that are executed at specific \textit{Phases}). We link the \textit{Tasks} to the \textit{Actors} and \textit{Information Exchange} they involve, to represent the stakeholder collaborations for each \textit{Task}. Finally, we relate the \textit{Actors} to their \textit{Roles} and \textit{Skills}, and also link the \textit{Roles} to the \textit{Skills} they require. 

Adding the property \textit{is missing} to any of the 6 elements in the communication model is of great interest for documenting the challenges mentioned in the interviews. We chose to represent the communication patterns using these 6 elements precisely because challenges arise if any of them are missing. For example, an \textit{Actor} may miss specific \textit{Skills}, or a fairness-related \textit{Task} may be entirely missing. 
 
Adding information on the affiliation of \textit{Actors} is also of interest to better describe the stakeholders involved in fairness-related decisions, and to identify potential issues with conflict of interests, privacy, accountability, or legal frameworks. 

The elements and properties of this conceptual framework were sufficient to represent the communication patterns we observed in the interviews. 
Adding more elements or properties would come at the risk of making it harder to generalize to new contexts. 

In the next section, we apply this conceptual framework to illustrate three relevant patterns observed in the challenges.

\subsubsection{Pattern 1: Actors with a developer role are predominant and miss guidance from domain experts.}

Several participants mentioned challenges with the \textit{involvement} and \textit{role} of stakeholders with domain expertise (Table~\ref{tab:main_challenges}).
Actors with a developer role are predominant, especially at the beginning of an algorithm's life-cycle, i.e., formulation, development, and evaluation phases (Figure~\ref{fig:mainrolexphase}).
Developers make decisions that seem technical but have crucial implications for fairness and public policy. 
Yet, actors with domain expertise may not be involved in guiding such seemingly technical decisions. 
Actors with technical skills become the main decision-makers, while they potentially miss domain expertise skills and stakeholders with the roles of advisor and policy-makers. 

Such technical decisions with fairness implications include, e.g., balancing a model's False Positive and False Negative rates, or fairness metrics based on these error rates. Domain expertise is needed to assess the practical implications of each type of error\footnote{For instance in punitive use cases \cite{rodolfa}, False Positives (e.g., accusing innocents of fraud) can be more problematic than False Negatives (e.g., undetected fraud). In assistive use cases, False Negatives (e.g., failing to help individuals in need) can be more problematic than False Positives (e.g., helping less vulnerable individuals).}. 

\begin{figure}[t!]
\centering
\includegraphics[width=9cm]{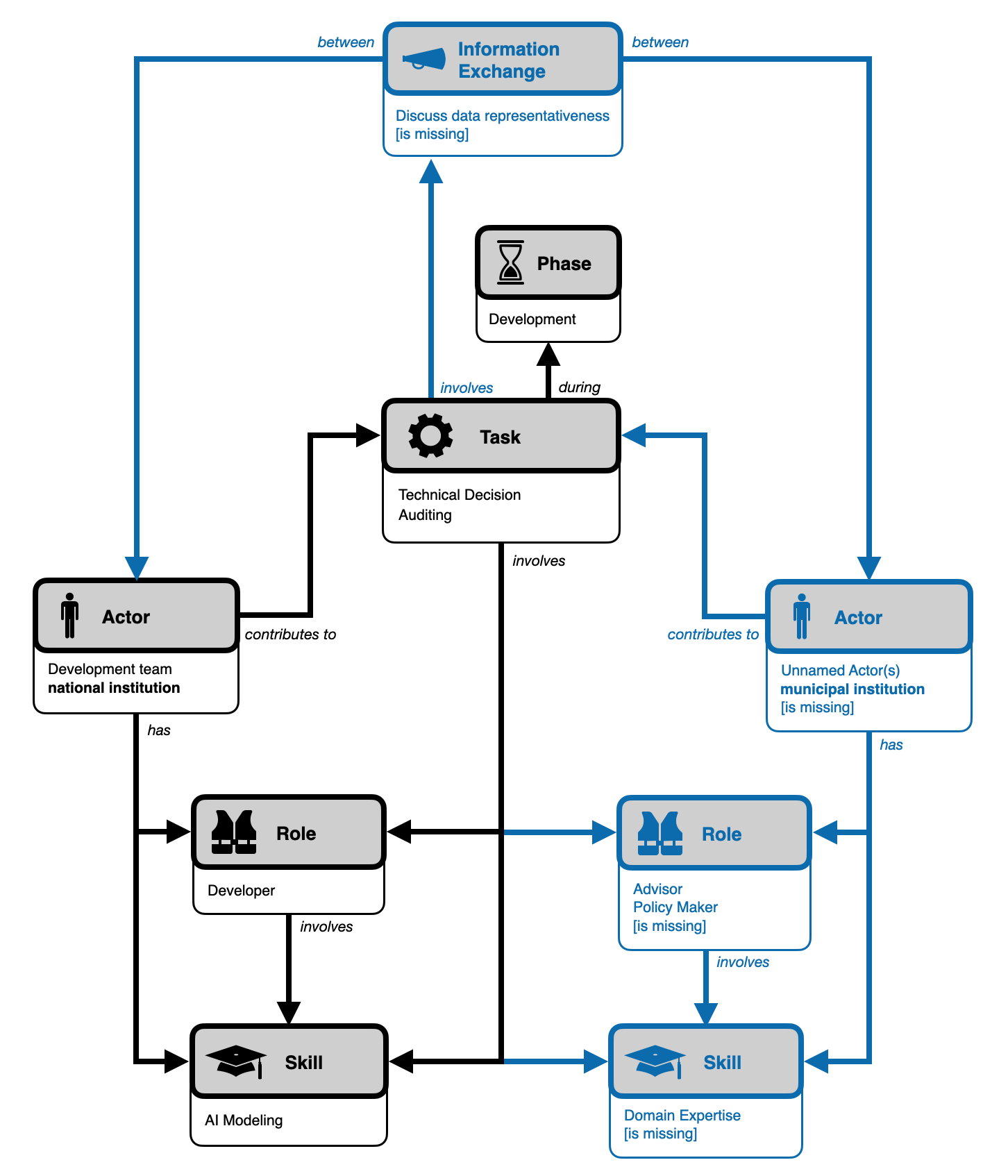}
\caption{Example of challenges with stakeholder \textit{involvement} and \textit{role} that constitute Pattern 1. Apparent technical decisions, such as defining which AI method to use and with which data features, have domain implications in practice but are made by developers alone. Other stakeholders with domain expertise are not involved in guiding the technical decisions. The missing information exchange is about the representativeness of the data features and their applicability to the use case.}
\label{fig:challenge_1}
\end{figure}

\begin{table}[h]
\centering
  \begin{tabular}{p{0.1\linewidth}p{0.8\linewidth}}
    \toprule
    Participant & Quote \\
    \midrule
    P2 & [For handling bias and error rates] \textit{"the technical colleagues give advice when the model is good enough, but it’s a bit of a grey area. We also rely on literature and on the technical teams’ judgment”.} \\
    P3 & \textit{“Hired an external bureau for auditing and investigating the algorithm”.} [Also because they] \textit{“couldn't get reliable predictions because the social domain changes all the time, and it’s hard to keep track of these changes”.}\\
    P5 & \textit{"An external company was hired to develop the model for the municipality, which made the data ecosystem quite complex''} \\ 
    P8 & \textit{"There should be more focus on asking users what policy-makers perceive as risks and biases” } \\
    P9 & \textit{"Involvement and direct information of the operators who work with AI system is needed, which particular change or improvement would be most useful for them"} \\
  \bottomrule
\end{tabular}
\vspace{5pt}
\caption{Quotes illustrating the communication pattern in Figure \ref{fig:challenge_1} where information on data quality was missing.}\label{tab:chall_finding1}
\end{table}

Our conceptual framework (Figure~\ref{fig:parts_model}) can be used to represent such communication issues. For example, Figure \ref{fig:challenge_1} illustrate the quotes from Table  \ref{tab:chall_finding1}. Developers must decide which data features are suitable for a use case, and how to use them withing AI systems. 
Domain experts could inform developers about the context in which the data features are representative of specific social groups. 
Without such information, developers may decide to use data features in ways that produce biased results for specific social groups. 

\subsubsection{Pattern 2: End-users and policy-makers may lack the technical skills to interpret the system’s limitations and uncertainty}

Several participants mentioned challenges with the \textit{interpretation} of a system's limitations and uncertainty  (Table~\ref{tab:main_challenges}).
At the deployment phase, actors with end-users and policy-maker roles may question whether the system delivers what it is supposed to perform, and how to interpret the validity of its results.
Yet, they may not have the technical skills to understand the limitations of a system. 
They may miss guidance from actors with a developer role, who have the technical skills to understand the uncertainty and the practical limitations they entail.

\begin{figure}[t!]
\centering
\includegraphics[width=9cm]{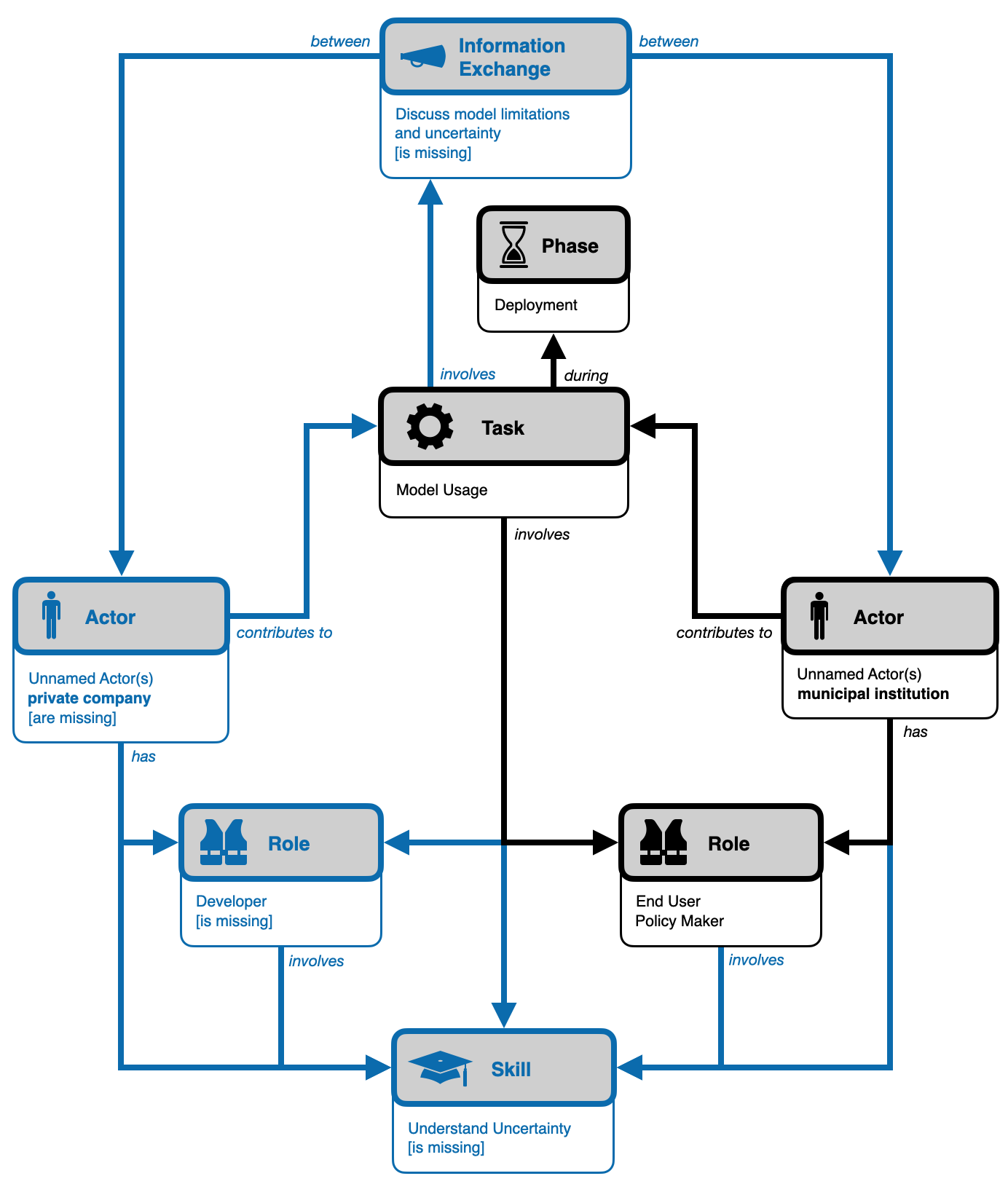}
\caption{Example of challenges with model \textit{interpretation}: the actors that use AI models, or make policies involving AI models, may miss the skills to understand model limitations and error metrics. They may also miss information exchange with developers who can explain the limitations and uncertainty. 
}
\label{fig:challenge_2}
\end{figure}

Our conceptual framework (Figure~\ref{fig:parts_model}) can be used to represent this communication challenge. For example, Figure \ref{fig:challenge_2} illustrates the quotes from Table \ref{tab:chall_finding2}. 

This finding highlights a need for increased input between end-users, policy-makers, and developers at the right phase. 
It echoes Pattern 1, where actors with a developer role lack guidance on the implications of their technical choices. These directly impact a system's limitations and uncertainty.

\begin{table}[t!]

  \begin{tabular}{p{0.1\linewidth}p{0.8\linewidth}}
    \toprule
    Participant & Quote \\
    \midrule
    P1 & \textit{Most important risk is that the model will not be used or is misinterpreted. For example, mixing up correlation and causality might lead to not helping people at risk of poverty.”} \\
    P3 & \textit{"We don’t know if governments and municipalities can understand the model.} \ 
    \tablefootnote{Upon asking their consent for publishing the quote, P3 added: \textit{"We don’t know if governments and municipalities [have the capacity in time and competence] to fully understand the model [so they can use it for policy tasks]."}} \\
   P5 & [The most difficult challenge is] \textit{“the gap between data scientists and policy-makers. How to make sure that what is developed is being well understood and useful for those of non-tech background.”} \\
   P6 & \textit{"Training for users is needed, to remind users not to rely on the tool but that the decision is up to them."} \\
    P8 & \textit{"The difficult for them to understand that there are many different interpretations. What it really means to be a ’true positive’, is this person really a fraud, or was this person not able to fill in the forms properly?”} \\
  \bottomrule
\end{tabular}
\vspace{3pt}
\caption{Quotes illustrating  the communication pattern in Figure \ref{fig:challenge_2}.
}
\label{tab:chall_finding2}
\end{table}

\begin{figure}[t!]
\centering
\includegraphics[width=9cm]{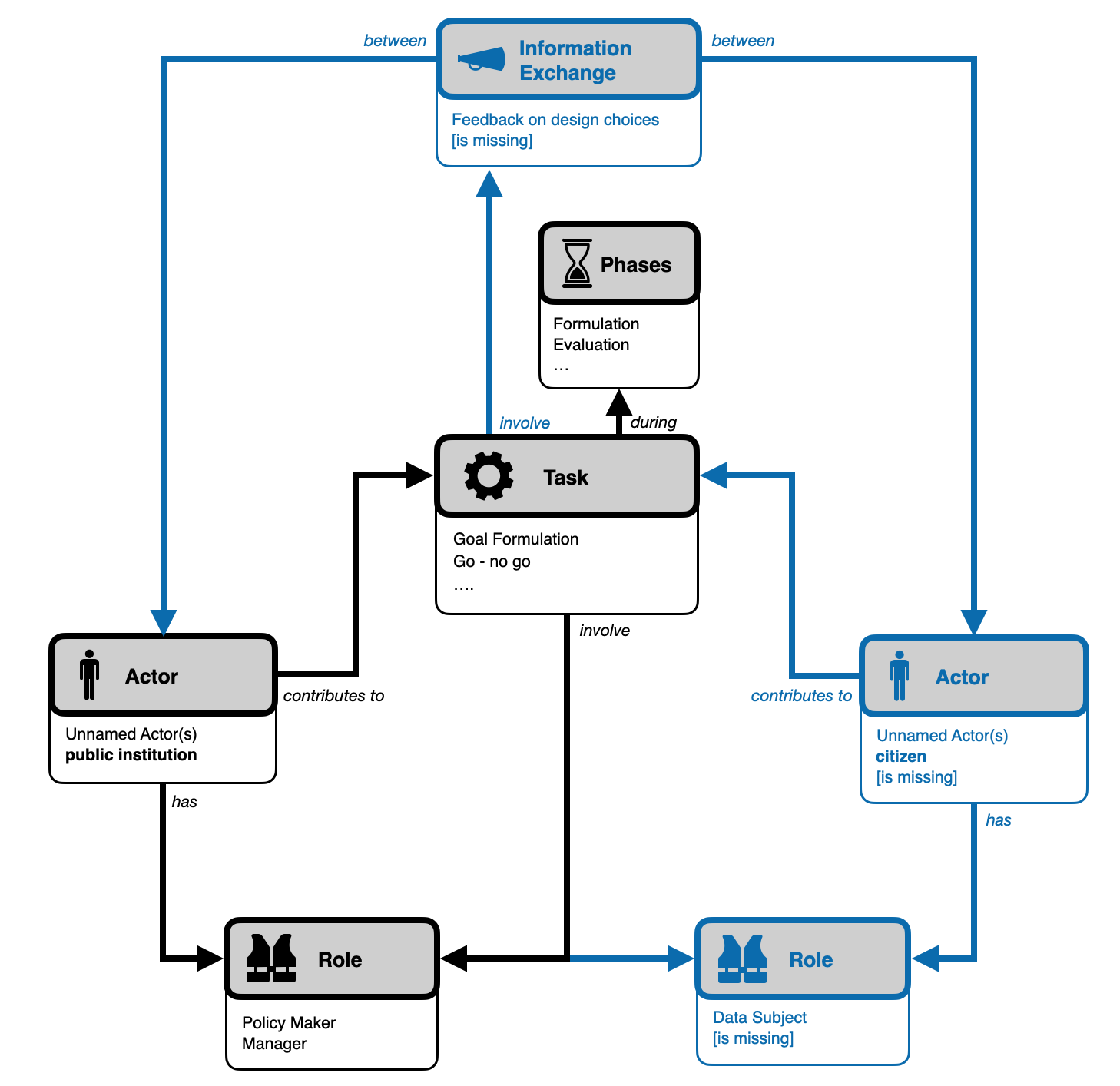}
\caption{Example of challenges with the \textit{involvement} of citizens: they are structurally absent throughout the algorithm's life-cycle, although they are the data subjects whose data is collected and processed, and who are impacted by the deployment of algorithmic systems. Information exchange is missing for them to understand and comment on the many design choices that impact fairness.}
\label{fig:challenge_3}
\end{figure}

\subsubsection{Pattern 3: Citizens are structurally absent throughout the algorithm's life-cycle. }

Some participants mentioned challenges with the lack of \textit{involvement} from citizens, e.g., who have the role of data subjects. 
For instance, in the formulation phase, citizen participation may be missing to give feedback on the design choices of the model. 
It is interesting to note that most participants did not mention citizens, and may thus overlook issues with their participation in a system's design or evaluation.
  
Our conceptual framework (Figure~\ref{fig:parts_model}) can be used to represent this communication challenge. For example, Figure \ref{fig:challenge_3} illustrates the quotes from Table \ref{tab:chall_finding3}. 
A lack of citizen involvement may lead to unbalanced fairness-related decisions that do not include key practical considerations. 

\begin{table}[t!]
  \begin{tabular}{p{0.1\linewidth}p{0.8\linewidth}}
    \toprule
    Participant & Quote \\
    \midrule
    P4 & \textit{``There is no direct citizen participation." } \\
    P5 & \textit{``it depends on the type of AI. If it has an impact on citizens or uses a lot of data from citizens, it would be relevant to include a focus group of citizens from the beginning but it is less relevant for e.g. road repairs.”}\\
    P7 & [On previous involvement with citizens] \textit{``they said no on the feasibility of the model from the municipality. They did not get it. It was more of a general no to technology instead of asking a targeted question”.}\\
  \bottomrule
\end{tabular}
\vspace{3pt}
\caption{
Quotes illustrating the communication pattern in Figure~\ref{fig:challenge_3}.
}
  \label{tab:chall_finding3}
\end{table}

\section{Discussion}\label{Discussion}
By characterizing the relations between the concepts in a conceptual framework, we demonstrated that unclear or undefined governance structures for roles, tasks, and skills can lead to misinterpretation of the system’s limitations and uncertainty, and even to misuses of the algorithmic system. 
From our use cases, we also saw that it was not always clear who makes final (mostly policy) decisions on the further development or use of algorithms, or what is the (legal, procedural, information) basis for such decisions. 
When there is a lack of actors filling the right roles at the right phase, actors can take on multiple roles at once for which they may not be fully equipped, which can lead to a discretionary imbalance.

It is possible that participants forgot to mention involved stakeholders in the development process or did not see some social participants as influential for choices, practices, and protocols. Forgetting a particular role or actor does not necessarily reflect the actual governance structure or experienced communication challenges. Participants may not have oversight, be unwilling to provide specific details, or were perhaps steered by how the interview questions were formulated. 

We also recognize that, in practice, formulations for roles and groups can vary and can be diffused. 
For instance, in some cases, actors identified as developers would primarily identify themselves as a researcher who would also carry out``developer tasks". 
We stress again that counting (co-)occurrences alone is not enough to assess the structure of the communication process that underlies fairness-related decisions. 
As mentioned in the results, sometimes an occurrence would be counted for role and phase when actually `role X was missing in phase Y', and thus the relations between concepts needed to be characterized to provide context to our findings. 
The conceptual framework we constructed covers comprehensively the relations that appear in the interviews, yet, it may not be sufficient for other scenarios. Fortunately, the incremental method applied for its construction allows for easy extension.

The number of interviews was limited \textit{(N=11)} and based on participants who collaborated on Dutch social domain use cases. 
The fact that all use cases reside in the Netherlands was for the sole reason that it was more readily available to us. 
It is important to emphasize that every (public sector) use case will have its own (normative) context, specific governance, and communication structure. 
With our use cases, we tried to appreciate these local conditions and resist “the portability trap” of stating that every AI use case will function the same from one context to another \cite{selbst2019fairness, barocas}. 

Regardless of the stated limitations, our findings confirm and complement earlier work emphasizing the growing autonomy and discretion of developers in public sector operations, as well as the unclear role divisions for the usage of automated decision tool  \cite{FestWieringa, Kalluri, Bovens2, Poel}. 
In terms of conceptual reorganization and synthesis, the number of interviews was sufficient to demonstrate the value of studying communication process underlying the choices and criteria for fairness-related decisions. Besides, the methodology we applied, consisting of (a) qualitative research, (b) qualitative coding analysis, (c) incremental construction of a conceptual framework, and (d) the application of the conceptual framework on acknowledged challenges, is rather generic, and we do not foresee constraints to its reuse in different, wider contexts. Yet, in terms of more general factual knowledge, 
More research is needed to investigate other local governance structures and communication processes around fairness-related decisions. 

\section{Conclusion}\label{conclusion}
In this research, we investigate fairness-related decisions through communication processes between diverse stakeholders that work on AI algorithms in the public sector.
We conducted semi-structured interviews to analyze the divisions of roles and tasks, the required skills, and the perceived challenges throughout the algorithm’s life-cycle. 
We applied qualitative coding analysis, to identify key elements of the communication processes that underlie fairness-related decisions. 
The results are formulated in a conceptual framework that represents these key elements as well as missing elements such as actors who miss skills or collaborators for certain tasks. To evaluate the adequacy and value of this methodology for the study of communication processes concerning fairness-related decisions, we applied it to social domain use cases in public organizations based in the Netherlands.  
The results we found are potentially relevant for policy interventions, as they generally indicate a lack of involvement and feedback between developer, end-user, and policy-maker roles. More precisely, we have captured the following key observations:
(i) Developers play the most prominent role in most tasks and phases of the system's life-cycle. They may miss guidance from stakeholders with advisor and policy-maker roles, and domain expertise skills. 
(ii) End-users and policy-makers often lack the technical skills to interpret the system's limitations and uncertainty and to estimate potential fairness issues. They rely on the technical skills of developers for making apparent technical decisions such as feature selection and balancing error rates, which potentially influence policy outcomes. 
(iii) Lastly, we observed that citizens are structurally absent throughout the system’s life-cycle, even though it is mentioned that their involvement is needed in the future for balanced fairness-related decision-making.  
These findings indicate that model governance is currently inadequate, and that there is a potential inability to recognize and address fairness issues throughout an algorithm's life-cycle. This can lead to misinterpretation and misuse of algorithms, with critical implications for the impacted populations. The conceptual framework we derived can help to address such issues before deployement, highlighting where to intervene (e.g. with adequate communications, gathering necessary skills currently missing, or introducing new roles), before the algorithm goes actually in production.



\backmatter


\begin{appendices}

\appendix

\begin{table}[h!]
  \caption{Questionnaire}
  \label{tab:structured_interviews}
  \begin{tabular}{p{0.7\linewidth}p{0.3\linewidth}}
    \toprule
    \textbf{Questions} & \textbf{Notes} \\
    \midrule
    Institution / Department &
  Name of entity/department \\
What is your (team's) role? &
  Description team/staff involved brief \\
Who do you work with (directly)? &
   \\
Domain and topic of use case &
   \\
Start and end date &
   \\
What type of system is being developed for the use case? &
  Intended use/aim \\
What is the goal of the system? &
   \\
External partners developing technology for the use case (if any) &
   \\
Who are the (end) users  - are they directly involved in the development process? &
   \\ \\
\textbf{II. Development process} &
   \\
Could you guide us through the process of development by mapping out phases - and specific actions in each phase? &
   \\
Could you guide us through the decisions made about the system and by whom? &
  e.g. involvement in deciding on: the goal of the system | design of the system (metrics/labeling/test/training/error) | evaluation of system | monitoring | deploying system \\
What kind of decisions do you and your team make? (could you give an example) &
   \\
What input is needed /do you use to make decisions as a reference point &
  e.g. handbook, training, expert group? \\
What kinds of exchanges are needed in your decision process? &
   \\
Are there other teams or (external) stakeholders involved in the decision-making process? &
   \\
How do you support the decision process of your collaborators with your output? &
   \\ \\
\textbf{III. Considerations} &
   \\
How can the development process be improved for the following, from your perspective? &
  e.g.,  information exchange, role division, handling error rates and biases, handling risks, responsibilities \\
What are the most difficult challenges and risks of failures for the system? &
   \\
How are these challenges and risks measured assessed and monitored? &
   \\
What information is needed (by whom) to handle these challenges and risks? &
   \\
Who is consulted for this information? &
   \\
What is your role in the process of addressing challenges and risks? &
   \\
Could any issues occur that might halt the development process, (if so could you give an example of how are these go/no-go decisions determined?) &
   \\
In real-life applications, could there be specific risks or negative impacts for individuals or social groups? &
   \\
Is error analysis / Bias analysis performed for negative impacts (if so how is this done and could you give an example)? &
   \\
Once the algorithm is deployed in practice, what kind of human oversight is available to control for error, bias, or negative impacts? &
   \\
What procedures and recourses, if any, are available for addressing the negative impacts of the system? &
   \\
Do you have access to explanations or training on the risks for individuals and social groups, e.g., from your colleagues or from external experts? &
   \\ \\
\textbf{IV. Follow-up questions} &
   \\
Who is in charge / responsible for mitigating measures on respecting privacy and data protection? For instance, is there a valid legal basis for processing personal data? &
   \\
Are there cybersecurity or privacy-preserving measures deployed to preserve privacy and data security? &
   \\
If no challenges (or very few) concerns are mentioned in 4, provide a scenario? &
  e.g. complaints about the output; security breaches; what if the training set is not representative,  high error rates
  \\
  \bottomrule
\end{tabular}
\end{table}

\end{appendices}

\bibliography{sn-bibliography}


\begin{thebibliography}{49}
\ifx \bisbn   \undefined \def \bisbn  #1{ISBN #1}\fi
\ifx \binits  \undefined \def \binits#1{#1}\fi
\ifx \bauthor  \undefined \def \bauthor#1{#1}\fi
\ifx \batitle  \undefined \def \batitle#1{#1}\fi
\ifx \bjtitle  \undefined \def \bjtitle#1{#1}\fi
\ifx \bvolume  \undefined \def \bvolume#1{\textbf{#1}}\fi
\ifx \byear  \undefined \def \byear#1{#1}\fi
\ifx \bissue  \undefined \def \bissue#1{#1}\fi
\ifx \bfpage  \undefined \def \bfpage#1{#1}\fi
\ifx \blpage  \undefined \def \blpage #1{#1}\fi
\ifx \burl  \undefined \def \burl#1{\textsf{#1}}\fi
\ifx \doiurl  \undefined \def \doiurl#1{\url{https://doi.org/#1}}\fi
\ifx \betal  \undefined \def \betal{\textit{et al.}}\fi
\ifx \binstitute  \undefined \def \binstitute#1{#1}\fi
\ifx \binstitutionaled  \undefined \def \binstitutionaled#1{#1}\fi
\ifx \bctitle  \undefined \def \bctitle#1{#1}\fi
\ifx \beditor  \undefined \def \beditor#1{#1}\fi
\ifx \bpublisher  \undefined \def \bpublisher#1{#1}\fi
\ifx \bbtitle  \undefined \def \bbtitle#1{#1}\fi
\ifx \bedition  \undefined \def \bedition#1{#1}\fi
\ifx \bseriesno  \undefined \def \bseriesno#1{#1}\fi
\ifx \blocation  \undefined \def \blocation#1{#1}\fi
\ifx \bsertitle  \undefined \def \bsertitle#1{#1}\fi
\ifx \bsnm \undefined \def \bsnm#1{#1}\fi
\ifx \bsuffix \undefined \def \bsuffix#1{#1}\fi
\ifx \bparticle \undefined \def \bparticle#1{#1}\fi
\ifx \barticle \undefined \def \barticle#1{#1}\fi
\bibcommenthead
\ifx \bconfdate \undefined \def \bconfdate #1{#1}\fi
\ifx \botherref \undefined \def \botherref #1{#1}\fi
\ifx \url \undefined \def \url#1{\textsf{#1}}\fi
\ifx \bchapter \undefined \def \bchapter#1{#1}\fi
\ifx \bbook \undefined \def \bbook#1{#1}\fi
\ifx \bcomment \undefined \def \bcomment#1{#1}\fi
\ifx \oauthor \undefined \def \oauthor#1{#1}\fi
\ifx \citeauthoryear \undefined \def \citeauthoryear#1{#1}\fi
\ifx \endbibitem  \undefined \def \endbibitem {}\fi
\ifx \bconflocation  \undefined \def \bconflocation#1{#1}\fi
\ifx \arxivurl  \undefined \def \arxivurl#1{\textsf{#1}}\fi
\csname PreBibitemsHook\endcsname

\bibitem[\protect\citeauthoryear{Rodolfa et~al.}{2021}]{rodolfa}
\begin{barticle}
\bauthor{\bsnm{Rodolfa}, \binits{K.T.}},
\bauthor{\bsnm{Lamba}, \binits{H.}},
\bauthor{\bsnm{Ghani}, \binits{R.}}:
\batitle{Empirical observation of negligible fairness--accuracy trade-offs in machine learning for public policy}.
\bjtitle{Nature Machine Intelligence}
\bvolume{3}(\bissue{10}),
\bfpage{896}--\blpage{904}
(\byear{2021})
\end{barticle}
\endbibitem

\bibitem[\protect\citeauthoryear{Rodolfa et~al.}{2020}]{DBLP:conf/fat/RodolfaSHMLG20}
\begin{bchapter}
\bauthor{\bsnm{Rodolfa}, \binits{K.T.}},
\bauthor{\bsnm{Salomon}, \binits{E.}},
\bauthor{\bsnm{Haynes}, \binits{L.}},
\bauthor{\bsnm{Mendieta}, \binits{I.H.}},
\bauthor{\bsnm{Larson}, \binits{J.}},
\bauthor{\bsnm{Ghani}, \binits{R.}}:
\bctitle{Case study: Predictive fairness to reduce misdemeanor recidivism through social service interventions}.
In: \bbtitle{Proceedings of the 2020 Conference on Fairness, Accountability, and Transparency}.
\bsertitle{FAT* '20},
pp. \bfpage{142}--\blpage{153}.
\bpublisher{Association for Computing Machinery},
\blocation{New York, NY, USA}
(\byear{2020}).
\doiurl{10.1145/3351095.3372863} .
\burl{https://doi.org/10.1145/3351095.3372863}
\end{bchapter}
\endbibitem

\bibitem[\protect\citeauthoryear{Williamson}{2016}]{williamsoneducation}
\begin{barticle}
\bauthor{\bsnm{Williamson}, \binits{B.}}:
\batitle{Digital education governance: data visualization, predictive analytics, and ‘real-time’policy instruments}.
\bjtitle{Journal of education policy}
\bvolume{31}(\bissue{2}),
\bfpage{123}--\blpage{141}
(\byear{2016})
\end{barticle}
\endbibitem

\bibitem[\protect\citeauthoryear{Van~Veenstra et~al.}{2019}]{TNO1}
\begin{botherref}
\oauthor{\bsnm{Van~Veenstra}, \binits{A.F.E.}},
\oauthor{\bsnm{Djafari}, \binits{S.}},
\oauthor{\bsnm{Grommé}, \binits{F.}},
\oauthor{\bsnm{Kotterink}, \binits{B.}},
\oauthor{\bsnm{Baartmans}, \binits{R.F.W.}}:
Quickscan AI in the Publieke dienstverlening
(2019).
\url{http://resolver.tudelft.nl/uuid:be7417ac-7829-454c-9eb8-687d89c92dce}
\end{botherref}
\endbibitem

\bibitem[\protect\citeauthoryear{Hoekstra et~al.}{2021}]{TNO2}
\begin{botherref}
\oauthor{\bsnm{Hoekstra}},
\oauthor{\bsnm{Chideock}},
\oauthor{\bsnm{Veenstra}, \binits{V.}}:
TNO Rapportage Quickscan AI in the Publieke sector II
(2021).
\url{https://www.rijksoverheid.nl/documenten/rapporten/2021/05/20/quickscan-ai-in-publieke-dienstverlening-ii}
\end{botherref}
\endbibitem

\bibitem[\protect\citeauthoryear{Mehrabi et~al.}{2021}]{mehrabi}
\begin{barticle}
\bauthor{\bsnm{Mehrabi}, \binits{N.}},
\bauthor{\bsnm{Morstatter}, \binits{F.}},
\bauthor{\bsnm{Saxena}, \binits{N.}},
\bauthor{\bsnm{Lerman}, \binits{K.}},
\bauthor{\bsnm{Galstyan}, \binits{A.}}:
\batitle{A survey on bias and fairness in machine learning}.
\bjtitle{ACM Computing Surveys (CSUR)}
\bvolume{54}(\bissue{6}),
\bfpage{1}--\blpage{35}
(\byear{2021})
\end{barticle}
\endbibitem

\bibitem[\protect\citeauthoryear{Fass et~al.}{2008}]{fassCOMPAS}
\begin{barticle}
\bauthor{\bsnm{Fass}, \binits{T.L.}},
\bauthor{\bsnm{Heilbrun}, \binits{K.}},
\bauthor{\bsnm{DeMatteo}, \binits{D.}},
\bauthor{\bsnm{Fretz}, \binits{R.}}:
\batitle{The lsi-r and the compas: Validation data on two risk-needs tools}.
\bjtitle{Criminal Justice and Behavior}
\bvolume{35}(\bissue{9}),
\bfpage{1095}--\blpage{1108}
(\byear{2008})
\end{barticle}
\endbibitem

\bibitem[\protect\citeauthoryear{Commission et~al.}{2020}]{EU1}
\begin{botherref}
\oauthor{\bsnm{Commission}, \binits{E.}},
\oauthor{\bsnm{Communications~Networks}, \binits{C.}},
\oauthor{\bsnm{Technology}}:
The Assessment List for Trustworthy Artificial Intelligence (ALTAI) for self assessment.
Publications Office
(2020).
\doiurl{10.2759/002360} .
\url{https://data.europa.eu/doi/10.2759/002360}
\end{botherref}
\endbibitem

\bibitem[\protect\citeauthoryear{European~Commission and Technology}{2019}]{EU2}
\begin{botherref}
\oauthor{\bsnm{European~Commission}, \binits{C.} \bsuffix{Directorate{-}General for Communications~Networks}},
\oauthor{\bsnm{Technology}}:
Ethics Guidelines for Trustworthy Artificial Intelligence.
Publications Office
(2019).
\doiurl{10.2759/346720} .
\url{https://data.europa.eu/doi/10.2759/346720}
\end{botherref}
\endbibitem

\bibitem[\protect\citeauthoryear{Commission}{2021}]{AIACT}
\begin{botherref}
\oauthor{\bsnm{Commission}, \binits{E.}}:
Proposal for a regulation of the European parliament and of the council: laying down harmonised rules on artificial intelligence (artificial intelligence act) and amending certain union legislative acts
(2021).
\url{https://eur-lex.europa.eu/legal-content/EN/TXT/?uri=celex\%3A52021PC0206}
\end{botherref}
\endbibitem

\bibitem[\protect\citeauthoryear{Suresh and Guttag}{2021}]{Suresh}
\begin{bchapter}
\bauthor{\bsnm{Suresh}, \binits{H.}},
\bauthor{\bsnm{Guttag}, \binits{J.V.}}:
\bctitle{A framework for understanding sources of harm throughout the machine learning life cycle}.
In: \bbtitle{{EAAMO} 2021: {ACM} Conference on Equity and Access in Algorithms, Mechanisms, and Optimization, Virtual Event, USA, October 5 - 9, 2021},
pp. \bfpage{17}--\blpage{1179}.
\bpublisher{{ACM}},
\blocation{New York, NY, USA}
(\byear{2021}).
\doiurl{10.1145/3465416.3483305} .
\burl{https://doi.org/10.1145/3465416.3483305}
\end{bchapter}
\endbibitem

\bibitem[\protect\citeauthoryear{Lee et~al.}{2019}]{lee}
\begin{barticle}
\bauthor{\bsnm{Lee}, \binits{M.K.}},
\bauthor{\bsnm{Kusbit}, \binits{D.}},
\bauthor{\bsnm{Kahng}, \binits{A.}},
\bauthor{\bsnm{Kim}, \binits{J.T.}},
\bauthor{\bsnm{Yuan}, \binits{X.}},
\bauthor{\bsnm{Chan}, \binits{A.}},
\bauthor{\bsnm{See}, \binits{D.}},
\bauthor{\bsnm{Noothigattu}, \binits{R.}},
\bauthor{\bsnm{Lee}, \binits{S.}},
\bauthor{\bsnm{Psomas}, \binits{A.}}, \betal:
\batitle{Webuildai: Participatory framework for algorithmic governance}.
\bjtitle{Proceedings of the ACM on Human-Computer Interaction}
\bvolume{3}(\bissue{CSCW}),
\bfpage{1}--\blpage{35}
(\byear{2019})
\end{barticle}
\endbibitem

\bibitem[\protect\citeauthoryear{Amershi et~al.}{2019}]{amershi}
\begin{bchapter}
\bauthor{\bsnm{Amershi}, \binits{S.}},
\bauthor{\bsnm{Begel}, \binits{A.}},
\bauthor{\bsnm{Bird}, \binits{C.}},
\bauthor{\bsnm{DeLine}, \binits{R.}},
\bauthor{\bsnm{Gall}, \binits{H.}},
\bauthor{\bsnm{Kamar}, \binits{E.}},
\bauthor{\bsnm{Nagappan}, \binits{N.}},
\bauthor{\bsnm{Nushi}, \binits{B.}},
\bauthor{\bsnm{Zimmermann}, \binits{T.}}:
\bctitle{Software engineering for machine learning: A case study}.
In: \bbtitle{2019 IEEE/ACM 41st International Conference on Software Engineering: Software Engineering in Practice (ICSE-SEIP)},
pp. \bfpage{291}--\blpage{300}
(\byear{2019}).
\bcomment{IEEE}
\end{bchapter}
\endbibitem

\bibitem[\protect\citeauthoryear{Haakman et~al.}{2020}]{haakman}
\begin{botherref}
\oauthor{\bsnm{Haakman}, \binits{M.}},
\oauthor{\bsnm{Cruz}, \binits{L.}},
\oauthor{\bsnm{Huijgens}, \binits{H.}},
\oauthor{\bsnm{Deursen}, \binits{A.}}:
Ai lifecycle models need to be revised. an exploratory study in fintech.
arXiv preprint arXiv:2010.02716
(2020)
\end{botherref}
\endbibitem

\bibitem[\protect\citeauthoryear{Barocas et~al.}{2019}]{barocas-hardt-narayanan}
\begin{bbook}
\bauthor{\bsnm{Barocas}, \binits{S.}},
\bauthor{\bsnm{Hardt}, \binits{M.}},
\bauthor{\bsnm{Narayanan}, \binits{A.}}:
\bbtitle{Fairness and Machine Learning: Limitations and Opportunities}.
\bpublisher{The MIT Press},
\blocation{Cambridge, Massachusetts}
(\byear{2019}).
\burl{http://www.fairmlbook.org}
\end{bbook}
\endbibitem

\bibitem[\protect\citeauthoryear{Saleiro et~al.}{2018}]{Saleiro}
\begin{botherref}
\oauthor{\bsnm{Saleiro}, \binits{P.}},
\oauthor{\bsnm{Kuester}, \binits{B.}},
\oauthor{\bsnm{Hinkson}, \binits{L.}},
\oauthor{\bsnm{London}, \binits{J.}},
\oauthor{\bsnm{Stevens}, \binits{A.}},
\oauthor{\bsnm{Anisfeld}, \binits{A.}},
\oauthor{\bsnm{Rodolfa}, \binits{K.T.}},
\oauthor{\bsnm{Ghani}, \binits{R.}}:
Aequitas: A Bias and Fairness Audit Toolkit.
arXiv
(2018).
\doiurl{10.48550/ARXIV.1811.05577} .
\url{https://arxiv.org/abs/1811.05577}
\end{botherref}
\endbibitem

\bibitem[\protect\citeauthoryear{Stapleton et~al.}{2022}]{stapleton}
\begin{bchapter}
\bauthor{\bsnm{Stapleton}, \binits{L.}},
\bauthor{\bsnm{Saxena}, \binits{D.}},
\bauthor{\bsnm{Kawakami}, \binits{A.}},
\bauthor{\bsnm{Nguyen}, \binits{T.}},
\bauthor{\bsnm{Ammitzb{\o}ll~Fl{\"u}gge}, \binits{A.}},
\bauthor{\bsnm{Eslami}, \binits{M.}},
\bauthor{\bsnm{Holten~M{\o}ller}, \binits{N.}},
\bauthor{\bsnm{Lee}, \binits{M.K.}},
\bauthor{\bsnm{Guha}, \binits{S.}},
\bauthor{\bsnm{Holstein}, \binits{K.}}, \betal:
\bctitle{Who has an interest in “public interest technology”?: Critical questions for working with local governments \& impacted communities}.
In: \bbtitle{Companion Publication of the 2022 Conference on Computer Supported Cooperative Work and Social Computing},
pp. \bfpage{282}--\blpage{286}
(\byear{2022})
\end{bchapter}
\endbibitem

\bibitem[\protect\citeauthoryear{Filgueiras}{2022}]{filgueiras}
\begin{barticle}
\bauthor{\bsnm{Filgueiras}, \binits{F.}}:
\batitle{New pythias of public administration: ambiguity and choice in ai systems as challenges for governance}.
\bjtitle{Ai \& Society}
\bvolume{37}(\bissue{4}),
\bfpage{1473}--\blpage{1486}
(\byear{2022})
\end{barticle}
\endbibitem

\bibitem[\protect\citeauthoryear{Madaio et~al.}{2022}]{Madaio}
\begin{barticle}
\bauthor{\bsnm{Madaio}, \binits{M.}},
\bauthor{\bsnm{Egede}, \binits{L.}},
\bauthor{\bsnm{Subramonyam}, \binits{H.}},
\bauthor{\bsnm{Wortman~Vaughan}, \binits{J.}},
\bauthor{\bsnm{Wallach}, \binits{H.}}:
\batitle{Assessing the fairness of ai systems: Ai practitioners' processes, challenges, and needs for support}.
\bjtitle{Proceedings of the ACM on Human-Computer Interaction}
\bvolume{6}(\bissue{CSCW1}),
\bfpage{1}--\blpage{26}
(\byear{2022})
\end{barticle}
\endbibitem

\bibitem[\protect\citeauthoryear{Fest et~al.}{2022}]{FestWieringa}
\begin{barticle}
\bauthor{\bsnm{Fest}, \binits{I.}},
\bauthor{\bsnm{Wieringa}, \binits{M.}},
\bauthor{\bsnm{Wagner}, \binits{B.}}:
\batitle{Paper vs. practice: How legal and ethical frameworks influence public sector data professionals in the netherlands}.
\bjtitle{Patterns}
\bvolume{3}(\bissue{10}),
\bfpage{100604}
(\byear{2022})
\end{barticle}
\endbibitem

\bibitem[\protect\citeauthoryear{Saldaña}{2013}]{Saldana}
\begin{bbook}
\bauthor{\bsnm{Saldaña}, \binits{J.}}:
\bbtitle{The Coding Manual for Qualitative Researchers}.
\bsertitle{International series of monographs on physics}.
\bpublisher{SAGE},
\blocation{California, USA}
(\byear{2013})
\end{bbook}
\endbibitem

\bibitem[\protect\citeauthoryear{Ropohl}{1999}]{Ropohl1999}
\begin{barticle}
\bauthor{\bsnm{Ropohl}, \binits{G.}}:
\batitle{Philosophy of socio-technical systems}.
\bjtitle{Society for Philosophy and Technology Quarterly Electronic Journal}
\bvolume{4}(\bissue{3}),
\bfpage{186}--\blpage{194}
(\byear{1999})
\end{barticle}
\endbibitem

\bibitem[\protect\citeauthoryear{Latour}{1999}]{Latour1999}
\begin{barticle}
\bauthor{\bsnm{Latour}, \binits{B.}}:
\batitle{On recalling ant}.
\bjtitle{The sociological review}
\bvolume{47}(\bissue{1\_suppl}),
\bfpage{15}--\blpage{25}
(\byear{1999})
\end{barticle}
\endbibitem

\bibitem[\protect\citeauthoryear{Latour}{1992}]{latour1992}
\begin{barticle}
\bauthor{\bsnm{Latour}, \binits{B.}}:
\batitle{Where are the missing masses? the sociology of a few mundane artifacts}.
\bjtitle{Shaping technology/building society: Studies in sociotechnical change}
\bvolume{1},
\bfpage{225}--\blpage{258}
(\byear{1992})
\end{barticle}
\endbibitem

\bibitem[\protect\citeauthoryear{Latour}{1994}]{latour1994}
\begin{botherref}
\oauthor{\bsnm{Latour}, \binits{B.}}:
On technical mediation.
Common knowledge
\textbf{3}(2)
(1994)
\end{botherref}
\endbibitem

\bibitem[\protect\citeauthoryear{Chopra and SIngh}{2018}]{Chopra}
\begin{bchapter}
\bauthor{\bsnm{Chopra}, \binits{A.K.}},
\bauthor{\bsnm{SIngh}, \binits{M.P.}}:
\bctitle{Sociotechnical systems and ethics in the large}.
In: \bbtitle{Proceedings of the 2018 AAAI/ACM Conference on AI, Ethics, and Society}.
\bsertitle{AIES '18},
pp. \bfpage{48}--\blpage{53}.
\bpublisher{Association for Computing Machinery},
\blocation{New York, NY, USA}
(\byear{2018}).
\doiurl{10.1145/3278721.3278740} .
\burl{https://doi.org/10.1145/3278721.3278740}
\end{bchapter}
\endbibitem

\bibitem[\protect\citeauthoryear{Dolata et~al.}{2022}]{dolataSTS}
\begin{barticle}
\bauthor{\bsnm{Dolata}, \binits{M.}},
\bauthor{\bsnm{Feuerriegel}, \binits{S.}},
\bauthor{\bsnm{Schwabe}, \binits{G.}}:
\batitle{A sociotechnical view of algorithmic fairness}.
\bjtitle{Information Systems Journal}
\bvolume{32}(\bissue{4}),
\bfpage{754}--\blpage{818}
(\byear{2022})
\end{barticle}
\endbibitem

\bibitem[\protect\citeauthoryear{Slota et~al.}{2021}]{slota}
\begin{botherref}
\oauthor{\bsnm{Slota}, \binits{S.C.}},
\oauthor{\bsnm{Fleischmann}, \binits{K.R.}},
\oauthor{\bsnm{Greenberg}, \binits{S.}},
\oauthor{\bsnm{Verma}, \binits{N.}},
\oauthor{\bsnm{Cummings}, \binits{B.}},
\oauthor{\bsnm{Li}, \binits{L.}},
\oauthor{\bsnm{Shenefiel}, \binits{C.}}:
Many hands make many fingers to point: challenges in creating accountable ai.
AI \& SOCIETY,
1--13
(2021)
\end{botherref}
\endbibitem

\bibitem[\protect\citeauthoryear{Poel and Royakkers}{2011}]{Poel}
\begin{bbook}
\bauthor{\bsnm{Poel}, \binits{v.d.}},
\bauthor{\bsnm{Royakkers}}:
\bbtitle{Ethics, Technology, and Engineering : an Introduction}.
\bpublisher{Wiley-Blackwell},
\blocation{United States}
(\byear{2011})
\end{bbook}
\endbibitem

\bibitem[\protect\citeauthoryear{Bovens and Zouridis}{2002}]{Bovens2}
\begin{barticle}
\bauthor{\bsnm{Bovens}, \binits{M.}},
\bauthor{\bsnm{Zouridis}, \binits{S.}}:
\batitle{From street-level to system-level bureaucracies: How information and communication technology is transforming administrative discretion and constitutional control}.
\bjtitle{Public Administration Review}
\bvolume{62}(\bissue{2}),
\bfpage{174}--\blpage{184}
(\byear{2002})
\doiurl{10.1111/0033-3352.00168}
{\href{https://arxiv.org/abs/https://onlinelibrary.wiley.com/doi/pdf/10.1111/0033-3352.00168}{{https://onlinelibrary.wiley.com/doi/pdf/10.1111/0033-3352.00168}}}
\end{barticle}
\endbibitem

\bibitem[\protect\citeauthoryear{Kalluri}{2020}]{Kalluri}
\begin{barticle}
\bauthor{\bsnm{Kalluri}, \binits{P.}}:
\batitle{Don't ask if artificial intelligence is good or fair, ask how it shifts power}.
\bjtitle{Nature}
\bvolume{583}(\bissue{7815}),
\bfpage{169}--\blpage{169}
(\byear{2020})
\doiurl{10.1038/d41586-020-02003-2}
\end{barticle}
\endbibitem

\bibitem[\protect\citeauthoryear{Danaher}{2016}]{Danaher}
\begin{barticle}
\bauthor{\bsnm{Danaher}, \binits{J.}}:
\batitle{The threat of algocracy: Reality, resistance and accommodation}.
\bjtitle{Philosophy \& Technology}
\bvolume{29}(\bissue{3}),
\bfpage{245}--\blpage{268}
(\byear{2016})
\end{barticle}
\endbibitem

\bibitem[\protect\citeauthoryear{Hickok}{2022}]{hickok}
\begin{botherref}
\oauthor{\bsnm{Hickok}, \binits{M.}}:
Public procurement of artificial intelligence systems: new risks and future proofing.
AI \& society,
1--15
(2022)
\end{botherref}
\endbibitem

\bibitem[\protect\citeauthoryear{Siffels et~al.}{2022}]{Siffels}
\begin{botherref}
\oauthor{\bsnm{Siffels}, \binits{L.}},
\oauthor{\bsnm{Berg}, \binits{D.}},
\oauthor{\bsnm{Sch{\"a}fer}, \binits{M.T.}},
\oauthor{\bsnm{Muis}, \binits{I.}}:
Public values and technological change: Mapping how municipalities grapple with data ethics.
New Perspectives in Critical Data Studies,
243
(2022)
\end{botherref}
\endbibitem

\bibitem[\protect\citeauthoryear{Jonk and Iren}{2021}]{Jonk}
\begin{bchapter}
\bauthor{\bsnm{Jonk}, \binits{E.}},
\bauthor{\bsnm{Iren}, \binits{D.}}:
\bctitle{Governance and communication of algorithmic decision making: A case study on public sector}.
In: \bbtitle{2021 IEEE 23rd Conference on Business Informatics (CBI)},
vol. \bseriesno{1},
pp. \bfpage{151}--\blpage{160}
(\byear{2021}).
\bcomment{IEEE}
\end{bchapter}
\endbibitem

\bibitem[\protect\citeauthoryear{Wieringa}{2020}]{Wieringa}
\begin{bchapter}
\bauthor{\bsnm{Wieringa}, \binits{M.}}:
\bctitle{What to account for when accounting for algorithms: A systematic literature review on algorithmic accountability}.
In: \bbtitle{Proceedings of the 2020 Conference on Fairness, Accountability, and Transparency}.
\bsertitle{FAT* '20},
pp. \bfpage{1}--\blpage{18}.
\bpublisher{Association for Computing Machinery},
\blocation{New York, NY, USA}
(\byear{2020}).
\doiurl{10.1145/3351095.3372833} .
\burl{https://doi.org/10.1145/3351095.3372833}
\end{bchapter}
\endbibitem

\bibitem[\protect\citeauthoryear{Bovens}{2007}]{Bovens}
\begin{bchapter}
\bauthor{\bsnm{Bovens}, \binits{M.}}:
\bctitle{182 public accountability}.
In: \bbtitle{The Oxford Handbook of Public Management}.
\bpublisher{Oxford University Press},
\blocation{Oxford, United Kingdom}
(\byear{2007}).
\doiurl{10.1093/oxfordhb/9780199226443.003.0009} .
\burl{https://doi.org/10.1093/oxfordhb/9780199226443.003.0009}
\end{bchapter}
\endbibitem

\bibitem[\protect\citeauthoryear{Spierings and van~der Waal}{2020}]{Spierings}
\begin{botherref}
\oauthor{\bsnm{Spierings}, \binits{J.}},
\oauthor{\bsnm{Waal}, \binits{S.}}:
Algoritme: de mens in de machine - Casusonderzoek naar de toepasbaarheid van richtlijnen voor algoritmen
(2020).
\url{https://waag.org/sites/waag/files/2020-05/Casusonderzoek_Richtlijnen_Algoritme_de_mens_in_de_machine.pdf}
\end{botherref}
\endbibitem

\bibitem[\protect\citeauthoryear{Cobbe et~al.}{2023}]{cobbe}
\begin{botherref}
\oauthor{\bsnm{Cobbe}, \binits{J.}},
\oauthor{\bsnm{Veale}, \binits{M.}},
\oauthor{\bsnm{Singh}, \binits{J.}}:
Understanding accountability in algorithmic supply chains.
arXiv preprint arXiv:2304.14749
(2023)
\end{botherref}
\endbibitem

\bibitem[\protect\citeauthoryear{Fujii}{2018}]{Interview1}
\begin{bbook}
\bauthor{\bsnm{Fujii}, \binits{L.A.}}:
\bbtitle{Interviewing in Social Science Research, A Relational Approach}.
\bpublisher{Routledge},
\blocation{New York, NY; Abingdon, Oxon}
(\byear{2018})
\end{bbook}
\endbibitem

\bibitem[\protect\citeauthoryear{Goede et~al.}{2019}]{Interview2}
\begin{bbook}
\bauthor{\bsnm{Goede}, \binits{D.}},
\bauthor{\bsnm{Bosma}},
\bauthor{\bsnm{Pallister-Wilkins}}:
\bbtitle{Secrecy and Methods in Security Research A Guide to Qualitative Fieldwork}.
\bpublisher{Routledge},
\blocation{New York, NY; Abingdon, Oxon}
(\byear{2019})
\end{bbook}
\endbibitem

\bibitem[\protect\citeauthoryear{Strauss}{1987}]{strauss}
\begin{bbook}
\bauthor{\bsnm{Strauss}, \binits{A.L.}}:
\bbtitle{Qualitative Analysis for Social Scientists}.
\bpublisher{Cambridge university press},
\blocation{Cambridge}
(\byear{1987})
\end{bbook}
\endbibitem

\bibitem[\protect\citeauthoryear{Noy and McGuinness}{2001}]{Noy}
\begin{botherref}
\oauthor{\bsnm{Noy}, \binits{N.}},
\oauthor{\bsnm{McGuinness}, \binits{B.}}:
Ontology development 101: A guide to creating your first ontology.
Stanford Knowledge Systems Laboratory
(2001).
\url{https://protege.stanford.edu/publications/ontology_development/ontology101.pdf}
\end{botherref}
\endbibitem

\bibitem[\protect\citeauthoryear{van Hage et~al.}{2011}]{SEM}
\begin{barticle}
\bauthor{\bsnm{Hage}, \binits{W.}},
\bauthor{\bsnm{Malaisé}, \binits{V.}},
\bauthor{\bsnm{Segers}, \binits{R.}},
\bauthor{\bsnm{Hollink}, \binits{L.}},
\bauthor{\bsnm{Schreiber}, \binits{G.}}:
\batitle{Design and use of the simple event model (sem)}.
\bjtitle{Web Semantics: Science, Services and Agents on the World Wide Web}
\bvolume{9},
\bfpage{128}--\blpage{136}
(\byear{2011})
\doiurl{10.1093/oxfordhb/9780199226443.003.0009}
\end{barticle}
\endbibitem

\bibitem[\protect\citeauthoryear{Golpayegani et~al.}{2022}]{Golpayegani}
\begin{bbook}
\bauthor{\bsnm{Golpayegani}, \binits{D.}},
\bauthor{\bsnm{Pandit}, \binits{H.J.}},
\bauthor{\bsnm{Lewis}, \binits{D.}}:
\bbtitle{AIRO: An Ontology for Representing AI Risks Based on the Proposed EU AI Act and ISO Risk Management Standards}
vol. \bseriesno{55},
p. \bfpage{51}
(\byear{2022}).
\bcomment{IOS Press}
\end{bbook}
\endbibitem

\bibitem[\protect\citeauthoryear{Franklin et~al.}{2022}]{FMOontology}
\begin{bchapter}
\bauthor{\bsnm{Franklin}, \binits{J.S.}},
\bauthor{\bsnm{Bhanot}, \binits{K.}},
\bauthor{\bsnm{Ghalwash}, \binits{M.}},
\bauthor{\bsnm{Bennett}, \binits{K.P.}},
\bauthor{\bsnm{McCusker}, \binits{J.}},
\bauthor{\bsnm{McGuinness}, \binits{D.L.}}:
\bctitle{An ontology for fairness metrics}.
In: \bbtitle{Proceedings of the 2022 AAAI/ACM Conference on AI, Ethics, and Society},
pp. \bfpage{265}--\blpage{275}
(\byear{2022})
\end{bchapter}
\endbibitem

\bibitem[\protect\citeauthoryear{Tamburri et~al.}{2020}]{TamburriSkill}
\begin{bchapter}
\bauthor{\bsnm{Tamburri}, \binits{D.A.}},
\bauthor{\bsnm{Van Den~Heuvel}, \binits{W.-J.}},
\bauthor{\bsnm{Garriga}, \binits{M.}}:
\bctitle{Dataops for societal intelligence: a data pipeline for labor market skills extraction and matching}.
In: \bbtitle{2020 IEEE 21st International Conference on Information Reuse and Integration for Data Science (IRI)},
pp. \bfpage{391}--\blpage{394}
(\byear{2020}).
\bcomment{IEEE}
\end{bchapter}
\endbibitem

\bibitem[\protect\citeauthoryear{Selbst et~al.}{2019}]{selbst2019fairness}
\begin{bchapter}
\bauthor{\bsnm{Selbst}, \binits{A.D.}},
\bauthor{\bsnm{Boyd}, \binits{D.}},
\bauthor{\bsnm{Friedler}, \binits{S.A.}},
\bauthor{\bsnm{Venkatasubramanian}, \binits{S.}},
\bauthor{\bsnm{Vertesi}, \binits{J.}}:
\bctitle{Fairness and abstraction in sociotechnical systems}.
In: \bbtitle{Proceedings of the Conference on Fairness, Accountability, and Transparency},
pp. \bfpage{59}--\blpage{68}
(\byear{2019})
\end{bchapter}
\endbibitem

\bibitem[\protect\citeauthoryear{Barocas et~al.}{2021}]{barocas}
\begin{bchapter}
\bauthor{\bsnm{Barocas}, \binits{S.}},
\bauthor{\bsnm{Guo}, \binits{A.}},
\bauthor{\bsnm{Kamar}, \binits{E.}},
\bauthor{\bsnm{Krones}, \binits{J.}},
\bauthor{\bsnm{Morris}, \binits{M.R.}},
\bauthor{\bsnm{Vaughan}, \binits{J.W.}},
\bauthor{\bsnm{Wadsworth}, \binits{W.D.}},
\bauthor{\bsnm{Wallach}, \binits{H.}}:
\bctitle{Designing disaggregated evaluations of ai systems: Choices, considerations, and tradeoffs}.
In: \bbtitle{Proceedings of the 2021 AAAI/ACM Conference on AI, Ethics, and Society},
pp. \bfpage{368}--\blpage{378}
(\byear{2021})
\end{bchapter}
\endbibitem

\end{thebibliography}

\end{document}